\documentclass[english,aps,nofootinbib]{revtex4-1}
\usepackage[T1]{fontenc} \usepackage[latin1]{inputenc}
\usepackage{graphicx} \usepackage{amssymb}

\makeatletter

\usepackage{amsmath, amsthm,color,psfrag}
\usepackage{graphicx}

\usepackage{vmargin}
\setmarginsrb{3cm}{2cm}{3cm}{3cm}{1cm}{1cm}{1cm}{1cm}
\usepackage{amsmath,bbm}
\usepackage{rotating}
\usepackage{psfrag}
\usepackage{fancyhdr}

\usepackage{babel}
\makeatother

\newcommand{\va}{\scriptscriptstyle}

\newcommand{\R}{\mathbb{R}}

\newcommand{\Q}{\mathrm{Q}}
\newcommand{\g}{\mathfrak{g}}
\newcommand{\su}{\mathfrak{su}}

\newcommand{\be}{\nopagebreak[3]\begin{equation}}
\newcommand{\ee}{\end{equation}}
\newcommand{\ba}{\nopagebreak[3]\begin{eqnarray}}
\newcommand{\ea}{\end{eqnarray}}
\newcommand{\n}{\nonumber}
\newcommand{\bee}{\nopagebreak[3]\begin{equation*}}
\newcommand{\eee}{\end{equation*}}
\newcommand{\baa}{\nopagebreak[3]\begin{eqnarray*}}
\newcommand{\eaa}{\end{eqnarray*}}
\newcommand{\la}{\label}

\definecolor{blue}{rgb}{0,0,1}
\definecolor{green}{rgb}{0,1,0}
\definecolor{red}{rgb}{1,0,0}
\definecolor{van}{rgb}{1,0,1}
\definecolor{al}{rgb}{1,1,0}

\DeclareFontFamily{U}{rsfs}{}         % Formal Script            %
\DeclareFontShape{U}{rsfs}{m}{n}{<5> rsfs5 <6><7> rsfs7          %
  <8><9><10><10.95><12><14.4><17.28><20.74><24.88> rsfs10}{}     %
\DeclareMathAlphabet{\mathfs}{U}{rsfs}{m}{n}                     %
\newcommand{\mfs}[1]{\mathfs {#1}}                               %
\newcommand{\sH}{{\mfs H}}

\begin{document}

\title{Canonical quantization of non-commutative holonomies in 2+1 loop quantum gravity}

\author{K. Noui}

\affiliation{Laboratoire de Math\'ematiques et Physique 
Th\'eorique, Parc de Grammont, Tours, France.}

\author{A. Perez, D. Pranzetti}

\affiliation{Centre de Physique Th\'{e}orique,
\thanks{ Unit\'{e} Mixte de Recherche (UMR 6207) du CNRS et des Universit\'{e}s
Aix-Marseille I, Aix-Marseille II, et du Sud Toulon-Var; laboratoire
affili\'{e} \`{a} la FRUMAM (FR 2291).%
}\\ Campus de Luminy, Marseille, France.}

\begin{abstract}
In this work we investigate the canonical quantization of 2+1
gravity with cosmological constant $\Lambda>0$ in the canonical
framework of loop quantum gravity. The unconstrained phase space
of gravity in 2+1 dimensions is coordinatized by an $SU(2)$
connection $A$ and the canonically conjugate triad field $e$. A
natural regularization of the constraints of 2+1 gravity can be
defined in terms of the holonomies of $A_{\pm}=A\pm\sqrt\Lambda
e$. As a first step towards the quantization of these constraints
we study the canonical quantization of the holonomy of the
connection $A_{\lambda}=A+\lambda e$ (for $\lambda \in \R$) on the
kinematical Hilbert space of loop quantum gravity. The holonomy
operator associated to a given path acts non trivially on spin
network links that are transversal to the path (a crossing).  We
provide an explicit construction of the quantum holonomy operator.
In particular, we exhibit a close relationship between the action
of the quantum holonomy at a crossing and Kauffman's $q$-deformed
crossing identity (with $q=\exp({\mathrm{i}\hbar{\lambda}}/2)$).
The crucial difference is that (being an operator acting on the
kinematical Hilbert space of LQG) the result is completely
described in terms of standard $SU(2)$ spin network states (in
contrast to $q$-deformed spin networks in Kauffman's identity). We
discuss the possible implications of our result.
\end{abstract}
\maketitle
\newpage{}

\section{Introduction}

The link between the Jones Polynomial, Chern-Simons theory and
quantum gravity in 2+1 dimensions with non vanishing cosmological
constant has been first shown by Witten in the seminal papers
\cite{Witten:1988hc}. First, he showed that 2+1 dimensional (first order) gravity
can be reformulated in terms of a Chern-Simons theory whose gauge algebra is
the isometry algebra of the local solutions of Einstein equations. Then, he proposed
a path integral quantization of the Chern-Simons theory with compact gauge Lie groups $G$.
In the case where $G=SU(2)$, this quantization is closely related to the quantization
of Euclidean gravity with a positive cosmological constant, which is the only situation
where the gauge group is compact. The work of Witten has opened an incredible rich new way
of understanding 3-manifolds and knots invariants because the expectation values of Wilson
loops observables in Chern-Simons theory has lead to a new covariant definition of the Jones
polynomials and its generalizations. 

After this result, it was precisely shown by Reshetikhin and Turaev \cite{RT} that quantum groups
play a central role in the construction of 3-manifolds invariants and knots polynomials. The construction
of the Turaev-Viro invariant is a very nice illustration of this fact \cite{Turaev:1992hq}. These invariants can 
be viewed as a $q$-deformed version of Ponzano and Regge amplitudes. Moreover,
the asymptotic of the vertex amplitudes (the quantum $6j$-symbol)
has been shown to be related to the action of $2+1$ gravity with
non vanishing cosmological constant in the WKB approximation
\cite{Mizoguchi:1991hk}.

All this, strongly motivates the idea that it should be possible to recover (in
the context of loop quantum gravity \cite{lqg}) the Turaev-Viro amplitudes as
the physical transition amplitudes of 2+1 gravity with non-vanishing
cosmological constant. This has been so far explicitly shown only in
the simpler case for pure gravity with vanishing cosmological constant
\cite{Noui:2004iy}. 

Can we find a clear-cut relationship between the Turaev-Viro
amplitudes and the transition amplitudes computed from the canonical
quantization of 2+1 gravity with non vanishing cosmological constant?
Using the so-called combinatorial quantization, developped in the compact case
in \cite{Alekseev:1994au} and then generalized in non-compact situations in
\cite{BNR} and \cite{Cat}, one shows how quantum groups appear in the canonical
quantization and therefore one makes a link between covariant and canonical quantizations
of gravity. However, quantum groups do not appear in this framework from a bottom-up approach
but they are putten by hand for purposes of regularization. The kinematical Hilbert space 
is finite dimensional and expressed already in terms of quantum groups. Physical states are
obtained solving the quantum constraints that reduce, in that case, to requiring invariance
under the quantum group adjoint action. { The combinatorial quantization is certainly one of the most
powerful canonical quantization of 2+1 dimensional gravity because it is, to our knowledge, the
only quantization scheme that leads to an explicit construction of the
physical Hilbert space for any topology of the space surface.}

Loop quantum gravity in 2+1 dimensions is another framework where it is possible to
address this question. { The advantage of working with loop quantum gravity instead of with
the combinatorial quantization is that it could help us understanding quantum gravity in four dimensions.} 
 As in the combinatorial quantization, we starts by quantizing
the unreduced phase space of the theory and then imposes the
constraints at the quantum level (Dirac recipe). But, contrary to the combinatorial
quantization (where the non-reduced phase space is finite dimensional),  
there is an infinite
number of degrees of freedom before imposing the constraints,  which in the case of $2+1$ gravity are
encoded in the infinitely many polymer-like excitations represented by
spin network states. In LQG it is natural to interpret the Turaev-Viro
invariant as transition amplitudes between arbitrary pairs of such
graph-based states. Now, if the previous statement makes
sense, the Turaev-Viro amplitudes would have to be related to the
kinematical states of the canonical theory, namely classical $SU(2)$
spin networks. In contrast the Turaev-Viro amplitudes are constructed
from the combinatorics of $q$-deformed spin networks \cite{KL}. This
would imply that the understanding of the relationship between the
Turaev-Viro invariants and quantum gravity requires the understanding of
the dynamical interplay between classical spin-network states and 
$q$-deformed amplitudes. We shall find here some indications about how
this relationship can arise.

Let us first briefly recall the canonical structure of (Riemannian)
gravity in 2+1 dimensions. The action of departure is
\[ S(A,e)=\int_{\mathcal{M}}\mathrm{tr}\left[e\wedge
F\left(A\right)\right]+\frac{\Lambda}{6}\mathrm{tr}\left[e\wedge
e\wedge e\right]\, ,\] where $\Lambda\ge 0$, $e$ is a cotriad
field, and $A$ is an $SU(2)$ connection.

Assuming that the space time manifold has topology $\mathcal{M}=\Sigma\times\R$, and, upon the standard 2+1 decomposition, the phase space of the theory is parametrized
by the pullback to $\Sigma$ of $\omega$ and $e$. In local coordinates
we can express them in terms of the 2-dimensional connection $A_{a}^{i}$
and the triad field $e_{a}^{i}$ where $a=1,2$ are space coordinate
indices and $i,j=1,2,3$ are $su(2)$ indices. The Poisson bracket
among these is given by\begin{equation}
\{ A_{a}^{i}\left(x\right),\, e_{b}^{j}\left(y\right)\} =\epsilon_{ab}\ \delta_{j}^{i}\delta^{(2)}\left(x,\, y\right)\label{poiss}\end{equation}
 where $\epsilon_{ab}$ is the 2d Levi-Civita tensor. The phase
 space variables are subjected to the first class local constraints
\be d_Ae=0\ \ \ \ {\rm and } \ \ \ F(A)+\Lambda\ e \wedge e=0
\label{constraints}\ee

The basic kinematical observables are given by the holonomy of the
connection and appropriately smeared functionals of the triad
field $e$. Quantization of these (unconstrained) observables leads
to an { irreducible} representation on a Hilbert space, the so-called
kinematical Hilbert space $\sH_k$, with a
diffeomorphism invariant inner product (see
\cite{Lewandowski:2005jk} and references therein): states in
$\sH_k$ are given by functionals $\Psi[A]$ of the
(generalized) connection $A$ which are square-integrable with
respect to a diff-invariant measure. The holonomy acts simply by
multiplication while $e$ acts as the derivative operator
$e_{a}^{i}=-\mathrm{i}\hbar\epsilon_{ab}\
\delta_{j}^{i}{\delta}/{\delta A_{b}^{j}}\,$ (more precisely, the objects that correspond to the field $e$
in loop quantum gravity are the flux operators associated to curves in $\Sigma$, see Section \ref{fluxx}).

%\subsection{Quantization of the Constraints}

Dynamics is defined by
imposing the quantum constraints (defined by the representation of
(\ref{constraints}) as self adjoint operators in $\sH_k$) on the
kinematical states. More precisely, the quantum
constraint-equations of 2+1 gravity with cosmological constant can
be written as \be
\mathcal{G}\left[\alpha\right]\triangleright\Psi=\int_{\Sigma}\mathrm{Tr}[\alpha{\mathrm{d}_{A}e}]\triangleright\Psi=0
\label{gauss}\ee and
\be
C_{\Lambda}\left[N\right]\triangleright\Psi=\int_{\Sigma}\mathrm{Tr}\left[N\left({
F(A)+\Lambda\, e\wedge e}\right)\right]\triangleright\Psi=0
\label{curvature}\ee
 for all $\alpha,N\in\mathcal{C}^{\infty}(\Sigma,su(2))$. The previous
equations are formal at this stage. The difficulty resides in the fact that
the constraints are non linear functional of the basic fields and their
quantization requires the introduction of a regularization.
Therefore, the precise meaning of the previous equations is a subtle
issue which will be at least partially investigated in this work.

In  \cite{Noui:2004iy} the quantization and solution
of the equations above for the special case $\Lambda=0$ is
completely worked out. More precisely, the construction of the
physical Hilbert space of 2+1 gravity is achieved by means of a
rigorous implementation of the Dirac quantization program to the
theory. A natural result of this work is the definition of the
path integral representation of the theory from the canonical
picture. This establishes the precise relationship between the
physical inner product of 2+1 gravity and the spin foam amplitudes
of the Ponzano-Regge model\footnote{See \cite{Bonzom-Freidel} for a more recent and alternative investigation of the link between the canonical quantization of the Wheeler-DeWitt equation and the symmetries of the Ponzano-Regge model.}. In addition to providing a systematic
definition of the quantum theory, the canonical treatment has the
advantage of automatically avoiding the infrared divergences that
plagued Ponzano-Regges original construction. Another advantage
of the formulation is that it sets the bases for the extension of
the analysis to the non vanishing cosmological constant
case\footnote{For a pedagogical review on the link between the
physical inner product and spin foams see \cite{Perez:2006gja}.
For more general basic literature about spin foams see
\cite{spinfoams}. {Recent results on the connection between LQG and spin foams in 4d can be found in \cite{SF-LQG}}.}. Indeed, the key observation is that equation
(\ref{curvature}) can be quantized by first introducing a
regulator consisting of a cellular decomposition $\Delta_{\Sigma}$
of $\Sigma$---with plaquettes $p\in\Delta_{\Sigma}$ of coordinate
area smaller or equal to $\epsilon^{2}$---so that\be
C_{0}\left(N\right)=\int_{\Sigma}\mathrm{Tr}\left[N\,
F\left(A\right)\right]=\lim_{\epsilon\rightarrow0}\sum_{p\in\Delta_{\Sigma}}\mathrm{Tr}\left[N_{p}\,
W_{p}\left(A\right)\right]\,, \label{qcurvature0}\ee
 where $W_{p}(A)=1+\epsilon^{2}F(A)+o(\epsilon^{2})\in SU(2)$ is
the Wilson loop computed in the fundamental representation. The quantization
of the previous expression is straightforward as the Wilson loop acts
simply by multiplication on the kinematical states of 2+1 gravity.
\begin{figure}[h]
\centerline{\hspace{0.5cm} \(
\begin{array}{c}
\includegraphics[width=10cm]{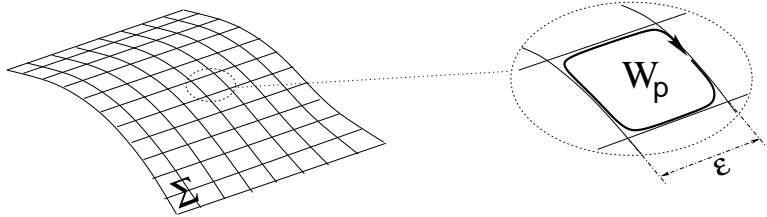}
\end{array}
\) }
%\ea
\caption{Cellular decomposition of the space manifold $\Sigma$ (a
square lattice in this example), and the infinitesimal plaquette
holonomy $W_p[A]$.} \label{regu}
\end{figure}
Then, the Ponzano-Regge amplitudes can be recovered through the definition of a physical scalar product by means of a projector operator into the kernel of (\ref{qcurvature0}). A key ingredient for this construction turns out to be, together with the background independence of the whole approach, the absence of anomaly in the quantum algebra of the constraints. In the case of $\Lambda\neq 0$, this is no longer the case, as shown in \cite{Anomaly} (see \cite{Ansatz} for a possible way around this difficulty).

Here, we propose an alternative approach to the problem of 2+1 gravity with $\Lambda\neq 0$ in the context of LQG. We start from the observation that, if we replace $W_{p}(A)$ by $W_{p}(A_{\pm})$ (with
$A_{\pm}=A\pm\sqrt{\Lambda} e$) on the previous equation, a simple
calculation shows that at the classical level we get\be
C_{\Lambda}\left[N\right]=\lim_{\epsilon\rightarrow0}\sum_{p\in\Delta_{\Sigma}}\mathrm{Tr}\left[N_{p}\,
W_{p}\left(A_{\pm}\right)\right]-\mathcal{G}\left[\pm\sqrt{\Lambda}N\right].
\label{qcurvature}\ee
 This provides a candidate background independent regularization of
the curvature constraint $C_{\Lambda}[N]$ for arbitrary values of
the cosmological constant. Notice that on gauge invariant states
(i.e. the solution space of the Gauss constraint) the second term
simply drops out. The quantization of the previous classical
expression requires the quantization of the holonomy of $A_{\pm}$.
More generally, as a first step towards the quantization of (\ref{qcurvature}), in the
present work we study the quantization of the holonomy $h_{\lambda}$ of the general
connection $A_{\lambda}\equiv A+\lambda e$ for $\lambda\in \R$. {The difficulties in the quantization of $h_{\lambda}$ arise from the fact that it is a {\it non-commutative} holonomy, since function of a connection ($A_{\lambda}$) becomes itself non-commutative upon quantization, as clear from the Poisson bracket (\ref{poiss}).}

The paper is
organized as follows: In Sections \ref{ns} we give a brief
account of our results avoiding technical details. In Section
 \ref{fluxx} we briefly recall the quantization scheme of the $e$-field in the LQG formalism. In Section
 \ref{quanti} the technical results are exhibit in detail. The
 crossing between quantum holonomies is defined in terms of a
 series expansion in powers of the cosmological constant. We prove
 that the series is well defined and can be summed to produce a
 simple result. However, the result depends on quantization
 choices. The choice of some natural prescription, such as the fully symmetrized ordering, yields unsatisfactory results, as shown in Section \ref{simord}. 
 In Section \ref{sec:duflo} we briefly introduce the Duflo isomorphism which provides a preferred quantization map in a given sense. 
 In Section \ref{nose},  we compute the action of the quantum holonomy defined by a suitable implementation of the  Duflo map in the LQG formalism.
 The action of an quantum holonomy on a transversal holonomy (both in the fundamental representation) exactly reproduces
Kauffmann's bracket. 
In Section \ref{discu}, we discuss the possible implications of our results in the framework of the question raised in this introduction.
 Some technical material is
presented in the Appendices.

%\section{Motivation}\label{moti}

\section{The results in a nut-shell}

\label{ns}

In this work we explore the quantization of the (one parameter
family of) classical (kinematical) observables\begin{equation}\label{qh}
h_{\eta}\left[A_{\lambda}\right]={P}\
\mathrm{e}^{-\int_{\eta}A+\lambda e}\end{equation}
associated with a path $\eta\in \Sigma$, as
operators on the kinematical Hilbert space of 2+1 loop quantum
gravity.

Due to the tensorial form of the Poisson bracket (\ref{poiss})
(inherited by the commutator in the quantum theory) the action of
(\ref{qh}) on the vacuum simply creates a Wilson line excitation,
i.e. it acts simply by multiplication by the holonomy of $A$ along the
path, namely\be
h_{\eta}\left[A_{\lambda}\right]\triangleright1=h_{\eta}\left[A\right]\label{vido}.\ee
This is because the $e$-operator in the argument of the path ordered
exponential in (\ref{qh}) acts as a derivative operator with respect
to the components of the connection that are transversal to the curve
(notice the presence of the $\epsilon_{ab}$ in the canonical
commutation relations (\ref{poiss})). The action of the holonomy of $A_{\lambda}$ is
therefore expected not to be trivial when the loop $\alpha$ in
(\ref{qh}) is self intersecting or when it acts on generic
spin-network states containing vertices on (or edges transversal to)
$\alpha$.

Therefore, the simplest non-trivial example is the action on a
transversal Wilson loop in the fundamental representation. We
define the quantization of (\ref{qh}) by quantizing each term in
the series expansion of (\ref{qh}) in powers of $\lambda$. Terms
of order $n$  have $n$ powers of the $e$ operators. The
quantization of these products becomes potentially ill-defined due
to factor ordering ambiguities (operators associated to $e$ are
non commuting in the quantum theory \cite{zapa}). 
%If one attempts to solve this ambiguity using some natural prescription such as the fully symetrized ordering, even though a closed formula for the quantum holonomy action can be obtained, the final result is not correct, as show in Section \ref{simord} (for technical details see Appendix \ref{so}).

The same kind of problem 
has been recently investigated in \cite{Sahlmann}, where the authors provided an new derivation of the expectation values of holonomies in Chern-Simons theory. In the analysis of \cite{Sahlmann}, the same sort of ordering ambiguities arises due to the replacement of holonomy functionals under the path integral with a complicated functional differential operator; the authors show that the expected result can be recovered once a mathematically preferred ordering, dictated by the Duflo isomorphism, is adopted\footnote{For another application of the Duflo map in the context of $2+1$ quantum gravity see also \cite{Majid}.}. Therefore, following the example of \cite{Sahlmann}, we will also make use of this mathematical insight, but in our case the Duflo map will not do the all job. In fact, since the ambiguities in the quantization of 
(\ref{qh}) arise due to the presence of non-linear terms in the $e$-field, a second piece of information has to be taken into account, namely the quantum action of flux operators in LQG. 
Combining these two elements leads to a well defined quantization
for each term in the perturbative expansion in $\lambda$.
Moreover, the series can be summed and the result can be expressed
in a closed form, leading to algebraic structures
remarkably equal to those appearing in Kauffman's
$q$-deformed spin networks.

More precisely, if we concentrate on a single intersection (a {\em
crossing}) between the path defining the holonomy of $A_{\lambda}$
and a transversal spin-network edge  in the fundamental representation $j=1/2$ we obtain
 \ba\la{kauf} \begin{array}{c}
\includegraphics[width=1cm]{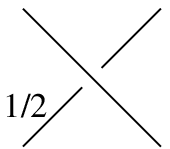}\end{array}=e^{\frac{i o\hbar \lambda}{4} }\begin{array}{c}
\includegraphics[width=1cm]{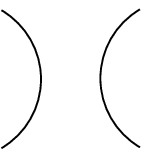}\end{array}
+e^{-\frac{i o\hbar \lambda}{4}  }\,
\begin{array}{c}
\includegraphics[width=1cm]{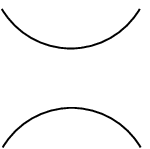}\end{array},
\ea 
where $o$ is the orientation of the crossing. Therefore, even though the crossing of paths
happens on the two dimensional manifold $\Sigma$, a distinction
between over and under crossing on the lhs of the previous expression is still possible according to the relative orientations of the path on which the quantum holonomy is defined and of the spin network edge it acts on. The action (\ref{kauf}) reproduce exactly Kaufmann's $q$-deformed crossing identity, where the deformation parameter reads $q=A^2=e^{\frac{i \hbar \lambda}{2} }$.

{ Despite of the strict resemblance of the previous equation and the Kauffman bracket, there objects appearing in equation (\ref{kauf})
are quite diffrerent from the ones in Kauffman's identity. Here, the paths involved are elements of $\sH_k$ of LQG, i.e. classical $SU(2)$ holonomies.
For that reason the famous Reidemeister identity as well as the Yang-Baxter braid identity
that can be derived from the analog of  (\ref{kauf}) in the knot theory context  are not valid here. Equation (\ref{kauf}) are  a different kind of
quantum deformation of the Maldestam relation for $SU(2)$ (the
binor spinorial identity) that we find using canonical
quantization of (\ref{qh}). This is the a central result of our
work.

The fact that our crossing does not satisfy the topological properties
of strands in knot theory deserves more qualification. As it is well known 2+1
gravity is a topological theory with no local degrees of freedom. In
the computation of expectation values of knotted (spacetime-embeded)
Wilson loops, this implies that their value is a knot-invariant as it
is shown in \cite{Witten:1988hc}.  From the viewpoint of the canonical
loop quantum gravity canonical approach
(where one reduces after quantization) this is expected to hold only on shell, i.e.,
after having imposed the quantum constraint (\ref{curvature}).
Our quantization of (\ref{qh}) is constructed at the level of the
kinematical Hilbert space where there are infinitely many local
(pure gauge) degrees of freedom. At that level there is no {\em a priori}
reason for the crossing to be topological.} { We will further discuss  this point in Section \ref{discu}.}

\section{Quantization of $e$-field}\label{fluxx}

In LQG there is a well-defined quantization of the $e$-field based
on the smearing of $e$ along one dimensional paths. More precisely,
given a path $\eta^a(t)\in \Sigma$ one considers the quantity \be
E(\eta)\equiv\int e_a^i \tau_i \frac{d\eta^a}{dt} dt=\int E^{ai}
\tau_i n_a dt, \ee where in the second equation we have replaced $e$
in terms of the connection conjugate momentum $E^a_i$ and $n_a\equiv
\epsilon_{ab}\frac{d\eta^a}{dt}$ is the normal to the path.
Therefore, the previous quantity represents the flux of $E$ across the
curve $\eta$.  The quantum operator associated to $E(\eta)$ acts
non trivially only on holonomies $h_{\gamma}$ along a path $\gamma\in
\Sigma$ that are transversal to $\eta$. It sufices to give its
action on trasnversal holonomies that either end or start on $\eta$.
The result is: 
\ba \label{flux}\hat E(\eta)\triangleright h_{\gamma}=\frac{1}{2}\hbar
\left\{\begin{array}{ccc} \!\! o(p)\tau^i\otimes \tau_i h_{\gamma}\ \ \mbox{if $\gamma$
ends at $\eta$}\\ o(p) h_{\gamma}\tau^i\otimes \tau_i\ \ \mbox{if $\gamma$ starts at
$\eta$}\end{array}\right., 
\ea
 where $o(p)$ is the orientation of
the intersection $p\in \Sigma$ (denoted $p$ for puncture), namely \be
o(p)=\left.\frac{\epsilon_{ab}\dot{\eta}^{a}
\dot{\gamma}^{b}}{\left|\epsilon_{ab}\dot{\eta}^{a}
\dot{\gamma}^{b}\right|}\right|_{p}\, \ee at the intersection $p\in \Sigma$.  In other
words the operator $E(\eta)$ acts at a puncture as an $SU(2)$
left-invariant-vector-field (LIV) if the puncture is the source of $h_{\gamma}$,
and it acts as a right-invariant-vector-field (RIV) if the puncture is the target
of $h_{\gamma}$. This observation will lead to a natural regularization of the
quantum holonomy operator (\ref{qh}) is what follows.

\section{Quantization of $h\left(A_{\lambda}\right)$}

\label{quanti}

Let $\Sigma\times\mathbb{R}$ be a global decomposition of the
$2+1$ dimensional spacetime,
$\gamma,\,\eta:\left(0,\,1\right)\rightarrow\Sigma$ two curves
that cross each other \emph{transversally} in
$\gamma\left(s_{*}\right)=\eta\left(t_{*}\right)$. Let
$A=A_{a}^{i}\mathrm{d}x^{a}\otimes\tau_{i}$ be a connection on a
principal $SU\left(2\right)$-bundle over $\Sigma\times\mathbb{R}$,
for which we choose a trivialization around
$\gamma\left(s_{*}\right)=\eta\left(t_{*}\right)$. Let
$h_{\gamma}\left(A\right)$ denote the holonomy of $\gamma$ in this
trivialization. Let $\left(A_{\lambda}\right)_{a}^{i}=A_{a}^{i}+
\lambda\, e_{a}^{i}=A_{a}^{i}+\lambda\,\epsilon_{ab}E_{i}^{b}$,
$E_{i}^{b}$ being the momentum canonically conjugate to
$A_{a}^{i}$.

Let us show that the action of
$h_{\eta}[A_{\lambda}]$ on the vacuum is trivial, namely \be
h_{\eta}[A_{\lambda}]|0\rangle=h_{\eta}[A]|0\rangle,\label{vido2}\ee
which is simply equivalent to equation (\ref{vido}) were we use
Dirac's bracket-notation for the vacuum whose wave functional
$\langle A |0\rangle=1$. The momenta $E_{i}^{b}$ are formally
quantized as $E_{i}^{b}\left(x\right)\mapsto
{-\mathrm{i}\hbar}{\delta}/{\delta A_{i}^{b}\left(x\right)}$. In order
to give a meaning to the quantum operator
$h_{\eta}\left(A_{\lambda}\right)$ we first develop its classical
expression in powers of $\lambda$ and obtain, for the generic $p$th order,

\ba \nonumber && \lambda^p \sum_{n\geq p}\,\sum_{m\geq
p}\,\left(-1\right)^{m+n}\,\sum_{1\leq k_{1}<\cdots<k_{p}\leq
n}\,\int_{0}^{1}\mathrm{d}t_{1}\cdots\int_{0}^{t_{n-1}}\mathrm{d}t_{n}\,\int_{0}^{1}\mathrm{d}s_{1}\cdots\int_{0}^{s_{m-1}}\mathrm{d}s_{m}\\
\nonumber &&
\left[A\left(\eta\left(t_{1}\right)\right)\cdots A\left(\eta\left(t_{k_1-1}\right)\right)\,E(\eta(t_{k_1}))\cdots
E(\eta(t_{k_p}))\,A\left(\eta\left(t_{k_p+1}\right)\right)\cdots
A\left(\eta\left(t_{n}\right)\right)\right]\,\left|0\right\rangle \,.
\label{nove}\ea As the commutator \be[E\left(\eta\left(t_{k}\right)\right)\, ,
A\left(\eta\left(t_{p}\right)\right) ]=0,\ee due to the fact that both fields in the commutator are pulled-back on the same curve, only the $p=0$ term of the
previous series survives when acting on the vacuum. Thus  (\ref{vido2}) follows.
The previous argument is formal: choosing a system of coordinates $\left(s,\, t\right)$ around
$\eta$ (which we suppose sufficiently small) in which $\eta$ be
represented by $\eta\left(t\right)=\left(0,\, t\right)$ we see
that
$\delta\left(\eta\left(t_{p}\right)-\eta\left(t_{k}\right)\right)=
\delta\left(\left(0,\, t_{p}\right)-\left(0,\,
t_{k}\right)\right)=\delta\left(0\right)\,\delta\left(t_{p}-t_{k}\right)$
is singular.  Nevertheless, a more careful treatment based on a suitable
regularization where the flux line is replaced by a flux tube
(defined by a smooth thickening of the path $\eta$) leads to the
same conclusion \cite{jaco} as our formal shortcut. 

\begin{figure}
\psfrag{a}{$\eta$}
\psfrag{b}{$\gamma$}
\includegraphics[width=3cm]{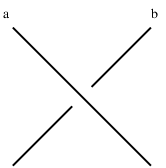}\ \ \ \ \ \ \ \ \ \ \ \ \ \ \ \ \ \ \ \ \ \ \ \ \ \includegraphics[width=3cm]{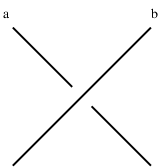}
\caption{Graphical representation of 
the action of two quantum holonomies $h_{\eta}(A_{\lambda})$ and $h_{\gamma}(A_{\lambda})$.
The three dimensional structure depicted as over-crossing or under crossing encodes operator ordering. In this way the picture on the left denotes the operator action
$h_{\eta}(A_{\lambda}) \triangleright h_{\gamma}(A_{\lambda})$ while the one on the right denotes $h_{\gamma}(A_{\lambda})\triangleright h_{\eta}(A_{\lambda})$. }\label{crossing}
\end{figure}

Let us move on now  and study
of the action of $\eta$ on $\gamma$. Denoting this action by
``$\triangleright$'' and using the previous results, we have:
\ba && \nonumber h_{\eta}\left(A_{\lambda}\right)
\triangleright h_{\gamma}\left(A_{\lambda}\right)\left|0\right\rangle
=h_{\eta}\left(A_{\lambda}\right)\triangleright
h_{\gamma}\left(A\right)\left|0\right\rangle =
\\ \nonumber &&
\left(1+\!\sum_{1\leq
n}\!(-1)^{n}\!\!\int_{0}^{1}\!\!\!\!\mathrm{d}t_{1}\cdot\cdot\!\!\int_{0}^{t_{n-1}}\!\!\!\!\!\!\!\!\!\!\!\mathrm{d}t_{n}\,
A_{\lambda}\left(\eta\left(t_{1}\right)\right)\cdot\cdot
A_{\lambda}\left(\eta\left(t_{n}\right)\right)\right) \triangleright\\
&&\ \ \ \ \ \ \ \ \ \ \ \ \ \ \ \ \ \ \ \ \ \left(1+\!\sum_{1\leq
m}\!(-1)^{m}\!\!\int_{0}^{1}\!\!\!\!\mathrm{d}s_{1}\cdot\cdot\!\!\int_{0}^{s_{m-1}}\!\!\!\!\!\!\!\!\!\!\!\mathrm{d}s_{m}\,
A\left(\gamma\left(s_{1}\right)\right)\cdot\cdot
A\left(\gamma\left(s_{m}\right)\right)\right)\left|0\right\rangle.\nonumber \ea

Developing in powers of $\lambda$ the coefficient at order $p$ is:
\ba \nonumber && \lambda^p\sum_{n\geq p}\,\sum_{m\geq
p}\,\left(-1\right)^{m+n}\,\sum_{1\leq k_{1}<\cdots<k_{p}\leq
n}\,\int_{0}^{1}\mathrm{d}t_{1}\cdots\int_{0}^{t_{n-1}}\mathrm{d}t_{n}\,\int_{0}^{1}\mathrm{d}s_{1}\cdots\int_{0}^{s_{m-1}}\mathrm{d}s_{m}\\
\nonumber &&
\left[A\left(\eta\left(t_{1}\right)\right)\cdots E(\eta(t_{k_1}))\cdots
E(\eta(t_{k_p}))\cdots
A\left(\eta\left(t_{n}\right)\right)\right]\,\triangleright
A\left(\gamma\left(s_{1}\right)\right)\cdots
A\left(\gamma\left(s_{m}\right)\right)\,.\ea 
In what follows we
shall omit the sums $\sum_{n\geq p},\,\sum_{m\geq p}$ and the
coefficient $\left(-1\right)^{m+n}$, and we shall only restore
them at the end of the calculations.
Let us concentrate on the action of the derivation operators on the connection
along $\gamma$. The relevant quantity is
\ba
\int_{0}^{1}\mathrm{d}s_{1}\cdots\int_{0}^{s_{m-1}}\!\!\!\! \mathrm{d}s_{m} \ E(\eta(t_{k_1}))\cdots
E(\eta(t_{k_p}))\,\triangleright
A\left(\gamma\left(s_{1}\right)\right)\cdots
A\left(\gamma\left(s_{m}\right)\right)\,.
\ea
One now uses
\ba \label{pipo} E(\eta(t))\,\triangleright
A\left(\gamma\left(s\right)\right)&=&\left(\epsilon_{ab}\dot{\gamma}^{a}\left(s_{*}\right)\dot{\eta}^{b}\left(t^{*} \right)\right)
\delta\left(\gamma\left(s\right)-\eta\left(t\right)\right)\nonumber \\ &=&
{\delta\left(s-s_{*}\right)\delta\left(t-t_{*}\right)}\frac{\epsilon_{ab}\dot{\gamma}^{a}\left(s_{*}\right)\dot{\eta}^{b}\left(t_{*}\right)}
{\left|\epsilon_{ab}\dot{\gamma}^{a}\left(s_{*}\right)\dot{\eta}^{b}\left(t_{*}\right)\right|}\nonumber \\ &=&
 o \ {\delta\left(s-s_{*}\right)\delta\left(t-t_{*}\right)},\ea
 where $o$ is the orientation of the intersection defined by taking $\gamma$ and $\eta$ in this order\,\footnote{There is an additional relative minus sign between under and over crossing. This can entirely encoded in $o$ if we choose the paths ordered according to the operator action (see Figure \ref{crossing}) and its caption.}. It is easy to see that only the terms containing $p$ consecutive graspings $E(\eta(t_q)), E(\eta(t_{q+1}))$ up to $E(\eta(q_{q+p}))$ which themselves act on $p$ consecutive $A(\gamma(s_k)), A(\gamma(s_{k+1}))$ up to $A(\gamma(s_{k+p}))$ 
survive. Any other possible term will vanish as a consequence of the previous equation (the domain of integration of the integrals of $A$'s evaluated on intermediate parameters will be constrained to a single point by the delta functions  (\ref{pipo})).
The Leibnitz rule now produces a sum over all possible orderings for the action of the $E$ on the sequence   $A(\gamma(s_k)), A(\gamma(s_{k+1}))$ up to $A(\gamma(s_{k+p}))$.
Finally, a factor $(1/p!)^2$ is produced by the ordered integral of $p$ two dimensional delta distributions\footnote{Here we are using that $$\int_{K}\delta(t_1) \cdots \delta(t_n) \, F(t_1,\cdots, t_n)=\frac{1}{p!} F(0,\cdots,0),$$ where $K=\left\{ t=\left(t_{1},\cdots,\,
t_{p}\right)\in\mathbb{R}^{p}\left|-\infty<t_{p}\leq\cdots\leq
t_{1}<\infty\right.\right\} $.}. One can arrange the integration variables and get  
  \ba\label{esa} &&\nonumber
\frac{\left(-\mathrm{i}o\hbar\lambda\right)^{p}}{p!}\,\sum_{k_{1}\geq1}\left(-1\right)^{k_{1}-1}\int_{t_{*}}^{1}\mathrm{d}t_{1}\cdots\int_{t_{*}}^{t_{k_{1}-2}}\mathrm{d}t_{k_{1}-1}\,
A\left(\eta\left(t_{1}\right)\right)\cdots
A\left(\eta\left(t_{k_{1}-1}\right)\right)\\ \nonumber &&
\tau^{i_{k_{1}}}\cdots\tau^{i_{k_{p}}}\,\sum_{v\geq0}\left(-1\right)^{v}\int_{0}^{t_{*}}\mathrm{d}\tilde{t}_{1}\cdots\int_{0}^{t_{v-1}}\mathrm{d}\tilde{t}_{v}\,
A\left(\eta\left(\tilde{t}_{1}\right)\right)\cdots
A\left(\eta\left(\tilde{t}_{v}\right)\right)\,\otimes\\ \nonumber &&
\,\sum_{\alpha_{k_{1}}\geq1}\left(-1\right)^{\alpha_{k_{1}}-1}\int_{s_{*}}^{1}\mathrm{d}s_{1}\cdots\int_{s_{*}}^{s_{\alpha_{k_{1}}-2}}\mathrm{d}s_{\alpha_{k_{1}}-1}\,
A\left(\gamma\left(s_{1}\right)\right)\cdots
A\left(\gamma\left(s_{\alpha_{k_{1}}-1}\right)\right)\\ &&
\tau_{(i_{k_{1}}}\cdots \tau_{i_{k_{p}})}\,\sum_{u\geq0}\left(-1\right)^{u}\int_{0}^{s_{*}}\mathrm{d}\tilde{s}_{1}\cdots\int_{0}^{s_{u-1}}\mathrm{d}\tilde{s}_{u}\,
A\left(\gamma\left(\tilde{s}_{1}\right)\right)\cdots
A\left(\gamma\left(\tilde{s}_{u}\right)\right)\,, \ea
where in the last line the brackets on the subindexes denote symmetrization, namely
\be \tau_{(i_{{1}}}\cdots \tau_{i_{{p}})}=\frac{1}{p!} \sum_{\pi\in S(p)}  \tau_{i_{{\pi(1)}}}\cdots \tau_{i_{{\pi(p)}}},\label{symm}\ee
for $S(p)$ denoting the group of permutations of $p$. The insertion of the symmetrized product of
generators can be thought of as the action of a quantization prescription defined by the map
\be\la{map1q}
\Q_S: ~E_{i_1}E_{i_2}\cdots E_{i_p}~~~\rightarrow~~~\frac{1}{p!}\sum_{\pi\in S(p)} \tau_{i_{\pi(1)}}\tau_{i_{\pi(2)}}\cdots\tau_{i_{\pi(p)}}.
\ee
As we have shown in the manipulations of this section, the previous quantization map arises naturally from the Leibnitz rule in our context.  
There are however factor ordering ambiguities due to the non-commutativity of the grasping operators that allow in principle for other prescriptions (that we will call $Q$ in the following section).
We will see in what follows that the advertised relationship with the Kauffman bracket is found if one uses 
the so-called Duflo map instead.

For further use it will be convenient to use the following graphical notation for the previous series
\be\label{zz}
\begin{array}{c}\psfrag{a}{$ $}\psfrag{b}{$ $}
\includegraphics[width=1cm]{x-quantum.eps}\end{array}=\begin{array}{c}\psfrag{a}{$Q_S$}
\includegraphics[width=1cm]{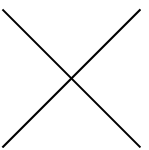}\end{array}+ z \begin{array}{c}
\includegraphics[width=1cm]{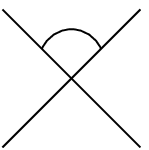}\end{array}+\frac{z^2}{2}\begin{array}{c}\psfrag{a}{\!$ \va Q_S$}
\includegraphics[width=1cm]{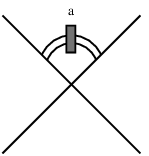}\end{array}+\frac{z^3}{3!}\begin{array}{c}\psfrag{a}{\!$\va Q_S$}
\includegraphics[width=1cm]{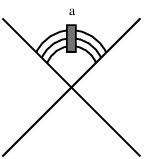}\end{array}+\cdots
\ee where $z=-io\hbar \lambda$, and  the boxes denote symmetrization (\ref{symm}) according to the quantization prescription $Q_S$ defined in (\ref{map1q}). 

\section{Summing up the perturbative series}\label{ambi}

In this Section we show that the perturbative expansion above can
be exactly summed once a definition of the symbol $Q$ is
provided. The completely symmetrized ordering $\Q\rightarrow \Q_S$
---which seems natural from the point of view of the Leibnitz rule
(see remark above)---leads to a complicated result.
A different crossing evaluation follows from the action (\ref{flux}) of the flux operator in LQG and the use of the Duflo isomorphism as a quantization map.
This possibility, which doesn't seem to contain any physical input but is mathematically preferred, as explained in more detail in the following, leads to the main result (\ref{kauf}) of this paper.

\subsection{Symmetric orderings}\la{simord}
The symmetric ordering, which we denote $\Q_S$,  arises naturally from the above treatment of the 
path ordered exponentials and the Leibnitz rule. As shown in Appendix \ref{so}, this prescription leads 
to a closed formula for the crossing, but it doesn't reproduce Kaufmann's bracket algebraic structure; namely, the fully symmetrized ordering yields
\be
\begin{array}{c}\psfrag{a}{}\psfrag{b}{}
\includegraphics{x-quantum.eps}\end{array}=B  \begin{array}{c}
\includegraphics{qL1.eps}\end{array}+C\begin{array}{c}
\includegraphics{qL2.eps}\end{array}
\label{kau1},\ee
where 
\bee
B(\lambda)=\sin[{\hbar \lambda/4}](\frac{2i}{3}-{\hbar
\lambda/4})+\cos[{\hbar \lambda/4}](1+\frac{i{\hbar
\lambda/4}}{3})
\eee
and
\bee
C(\lambda)=-\sin[{\hbar \lambda/4}](\frac{2i}{3}+{\hbar
\lambda/4})+\cos[{\hbar \lambda/4}](1-\frac{i{\hbar
\lambda/4}}{3}).
\eee

One can devise another natural quantization prescription by taking the flux quantization of fluxes of Section \ref{fluxx} as a guiding principle.
Accordingly, there is no quantization ambiguity for the zeroth and first order.  At second order the symmetric ordering studied above can be used. As shown Appendix \ref{so},
the result is proportional to the Casimir $E^2$. Therefore, the second order term is proportional to the zeroth order. We can define the third order as the result of the (unambiguous) action of a single flux $E$ on the second order.
This gives an iterative definition of all orders and produces a quantization prescription that coincides with $Q_S$ up to second order.
However, as the previous case also at  second order one departures from the Kauffman bracket expected result. We compute for completeness all orders in Appendix \ref{so},  the result is
\[B(\lambda)=\cos[\sqrt{3}\hbar
\lambda/4]-\frac{4i}{\sqrt{3}} \sin[{\sqrt{3}\hbar \lambda/4}]\]
and
\[C(\lambda)=\cos[\sqrt{3}\hbar
\lambda/4]+\frac{4i}{\sqrt{3}} \sin[{\sqrt{3}\hbar \lambda/4}].\]
This latter quantization prescription, has however, the advantage that all the ambiguities are now 
confined to the quantization of the Casimir $E^2$. The key ingredient in the resolution of this remaining ambiguity is the
existence of a preferred quantization prescription for Casimirs: the Duflo map. 

\subsection{The Duflo map}\la{sec:duflo}

%Let us now briefly describe the Duflo isomorphism, since it will provide us with a mathematically preferred ordering for the term $\Q(E_{i_{k_{1}}}\cdots E_{i_{k_{p}})}$ in (\ref{esa}). 

The Duflo map \cite{Duflo} is a generalization of the universal {\it quantization map} proposed by Harish-Chandra for semi-simple Lie algebras. 
The latter provides a prescription to quantize polynomials of commuting variables (the classical triad fields $e$) which after quantization acquire Lie algebra commutation relations (the flux operators $\hat E$). More precisely, given a set of commuting variables $E_i$ on the dual space $\g^*$ of the algebra $\g$, they generate the commutative algebra of polynomials, called the {\it symmetric algebra} over $\g$ and denoted Sym($\g$). If now we want to map this algebra into the one generated by non-commutative variables $\tau_i$ which satisfy the commutation relations $[\tau_i,\tau_j]=f_{ij}\,^k \tau_k$, we run into ordering problem since the commutative algebra Sym($\g$) must be mapped to the non-commutative {\it universal enveloping algebra} $U(\g)$. A natural quantization map introduced by Harish-Chandra \cite{Alekseev} is the so-called symmetric quantization, defined by its action on monomials, namely
\be\la{map1}
\Q_S: ~E_{i_1}E_{i_2}\cdots E_{i_n}~~~\rightarrow~~~\frac{1}{n!}\sum_{\pi\in S_n} \tau_{i_{\pi(1)}}\tau_{i_{\pi(2)}}\cdots\tau_{i_{\pi(n)}}.
\ee
A generalization of the previous map was provided by Duflo by composing it with a differential operator $j^{\frac{1}{2}}(\partial)$ on Sym($\g$), where $\partial_i\equiv\partial/\partial E_i$ represents derivatives with respect to the generators of Sym($\g$). In the case of the Lie algebra $\su(2)$, the Duflo map $\Q_D$ reads
\be\la{map2}
\Q_D=\Q_S\circ j^{\frac{1}{2}}(\partial)=\Q_S\circ\left(1+\frac{1}{12}\partial_i\partial_i+\cdots\right),
\ee
where the dots stand for terms containing higher derivatives.

The main property of $\Q_D$ is that given two Casimir elements A and B, the product of quantizations $\Q_D(A)\Q_D(B)$ coincides with the quantization of the product, $\Q_D(AB)$. Therefore, the Duflo map is an isomorphism between the invariant (under the action of $G$) sub-algebras Sym$(\g)^{\g}$ and $U(\g)^{\g}$. 

The Duflo map provides a mathematically preferred quantization for products of $E$; however, such choice is not always 
physically acceptable. For instance if one would use it for the quantization of angular momentum in the hydrogen atom one would get an energy spectrum incompatible with observations.
In LQG this map has also been proposed to provide an alternative quantization of the area operators \cite{Alekseev}. Such choice leads to a simpler area spectrum; however, it has drawback of violating  
cylindrical consistency \cite{alex}.

\subsection{Quantization in terms of flux operators}\label{nose}

In order to get the general form of the series (\ref{zz}) in the
case where we use the quantization of the flux operators given in
Section \ref{fluxx} it suffices to write the first few terms. In
the first order term, $E$ acts as a LIV on the portion of the
holonomy which has the crossing as its source and as a RIV on the
other one. The full result is, just as in (\ref{zz}): \be
\begin{array}{c}
\includegraphics[width=1cm]{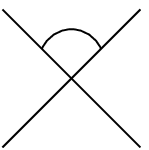}\end{array}
\ee 
In the second order diagram we have the action of two flux operators at the same point and therefore ordering ambiguities arise. In order to deal with them, we now use the prescription induced by the Duflo map, namely we write $(\tau_j\tau_k)$ as
\ba\la{tautau}
\Q_D[E_jE_k]&=&\Q_S\circ\left(1+\frac{1}{12}\partial_i\partial_i+\cdots\right)[E_jE_k]\n\\
&=&\frac{1}{2}(\tau_j\tau_k+\tau_k\tau_j)+\frac{1}{6}\delta_{jk}.
\ea
Diagrammatically, for the second order term we have
 \be
\begin{array}{c}\psfrag{a}{\!$\va Q_{\va D}$}
\includegraphics[width=1cm]{x2.eps}\end{array}=\frac{1}{2}\begin{array}{c}
\includegraphics[width=1cm]{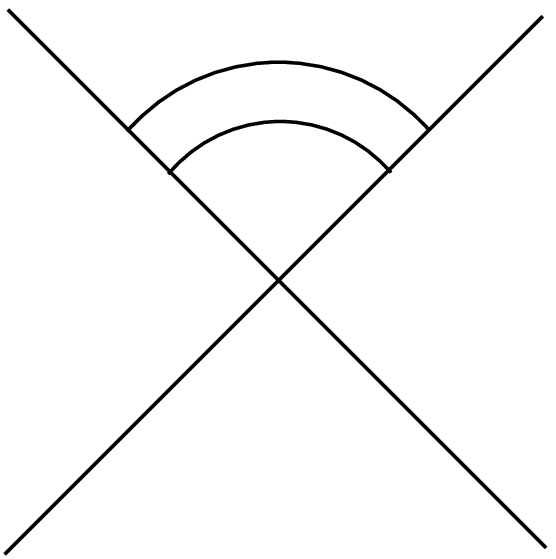}\end{array}+\frac{1}{2}\begin{array}{c}
\includegraphics[width=1cm]{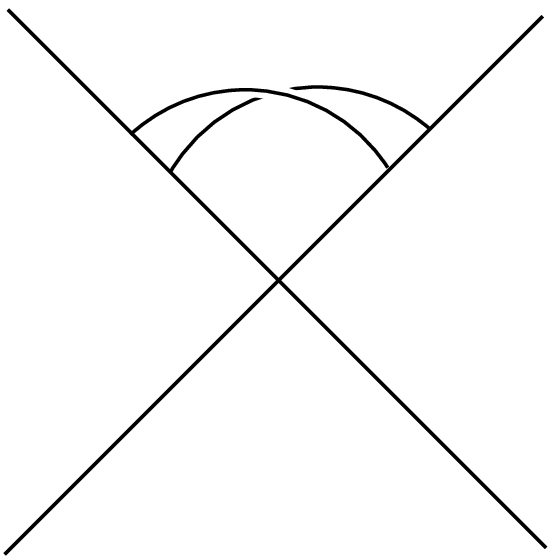}\end{array}+\frac{1}{6}\begin{array}{c}
\includegraphics[width=1cm]{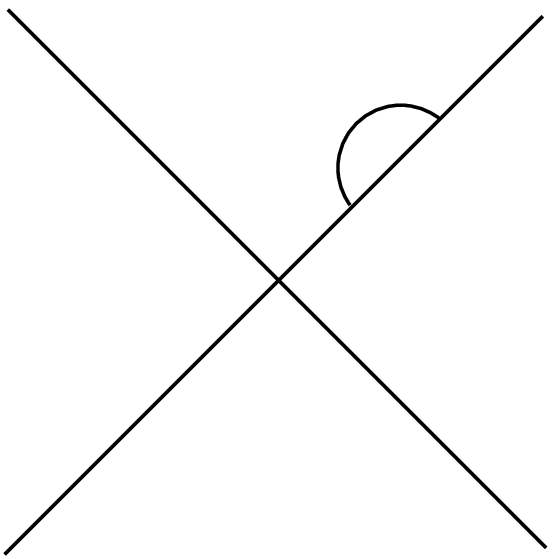}\end{array}=
\frac{1}{16}\begin{array}{c}
\includegraphics[width=1cm]{x.eps}\end{array},
\ee where in the second equality we used
the fact that $\{\tau^i,\tau^j\}=-1/2 \ \delta^{ij}$ and the value
of the Casimir in the fundamental representation. Therefore,
the second order diagram is proportional to the order zero
diagram. The third order
term is consequently proportional to the first order one and so
on \footnote{{Notice that, if we haven't used the quantization scheme of the flux operators proper of the LQG formalism, in order to compute the terms beyond the second order, we should have applied the Duflo map at all orders (i.e. compute the action of $Q_D$ on all the other products of $E$s). 
This alternative prescription (besides being much more involved) would lead to a result differing from the reproduction of the Kaufmann bracket, thus showing the central role played by the LQG representation of the fundamental variables.}}. We get in this way the general expression for arbitrary order.
Finally, choosing an orientation and using equations (\ref{bibi}) and (\ref{tauto}) in the Appendix \ref{app1}, we can express the ordered version of 
Equation (\ref{esa}) as
 \ba
 h_{\eta}\left(A_{\lambda}\right)\triangleright
h_{\gamma}\left(A_{\lambda}\right)\,\left|0\right\rangle
=\!\begin{array}{c}\psfrag{a}{}\psfrag{b}{}
\includegraphics{x-quantum.eps}\end{array}\!=\sum_{n\geq 0}\,\frac{(-z)^{n}}{4^n(n)!}
\begin{array}{c}
\includegraphics{qL1.eps}\end{array}
-\sum_{n\geq 0}\,\frac{(z)^{n}}{4^n(n)!}\begin{array}{c}
\includegraphics{qL2.eps}\end{array}.\n\\
\ea
Therefore, the series expansion in powers of $\lambda$ converges and leads to a simple
expression for the crossing.
Using Penrose convention
$\epsilon_{AB}\rightarrow i \epsilon_{AB}$ and $\epsilon^{AB}\rightarrow i
\epsilon^{AB}$ to take care of the different relative signs, the result is
\be
\begin{array}{c}\psfrag{a}{}\psfrag{b}{}
\includegraphics{x-quantum.eps}\end{array}=A  \begin{array}{c}
\includegraphics{qL1.eps}\end{array}+A^{-1}\begin{array}{c}
\includegraphics{qL2.eps}\end{array}
\label{kau},\ee
where $A=e^{\frac{i o\hbar \lambda}{4} }$, with $o$ the relative orientation between $\eta$ and $\gamma$. Equations (\ref{kau})
has the same form as Kauffman's $q$-deformed binor identity
for $q=\exp{i\lambda/2}$.

\section{Discussion}\label{discu}

We have shown that the holonomy of $A_\lambda$ in the fundamental representation
can be quantized in different ways due to ordering ambiguities. However, there { exists} a simple and 
natural quantization based on the Duflo map leading to the Kauffman-like algebraic structure 
for the action of the quantum holonomy defining a crossing. This result is { promissing in the road to finding} 
a relationship between Turaev-Viro amplitudes and physical amplitudes in canonical LQG.

%Even when we do not introduce a quantum group at any stage, and no
%dynamical constraint has been yet imposed, amplitudes such as the
%value of the quantum dimension (or self linking number of a Wilson
%loop in the language of \cite{Witten:1988hc}) $d_{q}=-q-q^{-1}$
%and $q=A^2=\exp{i\hbar\lambda/4}$ naturally appear from our
%treatment. Recall that the value of $d_q$ together with the deformed binor identity are the two ingredients for the
%combinatorial definition of the Turaev-Viro invariant according to
%the formulation of \cite{KL}. This is encouraging as it indicates
%that perhaps a strict correspondence between LQG and the
%Tuarev-Viro invariant can be established if one appropriately
%implements the next step: quantizing and imposing the curvature
%constraint (\ref{curvature}). This will be investigated in the future.

{The recovering of the Kauffman bracket related to the $q$-deformed crossing identity
is a remarkable result since it was obtained starting from the standard $SU(2)$ kinematical Hilbert space of LQG and 
combining the flux operators representation of the theory together with a mathematical input coming from the Duflo isomorphism. 
The fact that the crossing of our quantum holonomies have this structure is an encouraging result in finding a link between the role of quantum groups 
in 3d gravity with non vanishing cosmological constant and its canonical quantization.  However, the full link can only be established if  the dynamical  input from the implementation of the constraints
(\ref{curvature}) is brought in. Quantum holonomies defined here might be the right tool for regularizing the quantum constraints as proposed in (\ref{qcurvature}).

As pointed out  in the previous paragraph and at the end of Section II, the topological features of knots (Reidermeister moves) as well as the related quantum evaluation of Wilson loops is only to be found through dynamical considerations.
  Since in the present analysis no quantum group structure has been introduced by hand at any stage, at the present kinematical level, loops still evaluate according to the classical $SU(2)$ recoupling theory. Nevertheless, an intriguing indication  that the implementation of dynamics could lead to the emergence of the quantum dimension for loops evaluation is available already at this stage. More precisely, if one takes seriously the expression (\ref{qcurvature}) as a proposal for the regularized version of the curvature constraint (\ref{curvature})---notice that, in the naive continuum limit, the expression (\ref{curvature}) is recovered---then one could compute it's algebra by studying the action of the commutator on some states. The classical constraint algebra dictates that this should be  proportional to the Gauss constraint. If one performs this analysis, it is immediate to see that there are two types of anomalous contributions: one of the same kind of the anomaly  found in \cite{Anomaly}  (which could be called mild as the terms produced vanish when  acting  on gauge invariant states), and  another anomalous contribution (a stronger one) that does not annihilate  gauge invariant states. The latter anomalous terms happen to be proportional $(A^2+A^{-2}+\begin{array}{c}  \includegraphics[width=0.4cm,angle=360]{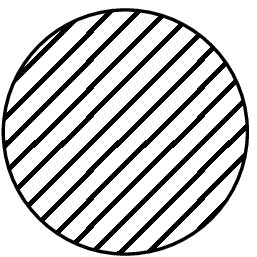}\end{array})$, where $\begin{array}{c}  \includegraphics[width=0.4cm,angle=360]{0loop.eps}\end{array}$ represents the loop with no area in the fundamental representation $j=1/2$. Thus the condition that an infinitesimal loop evaluates to the quantum dimension $-A^2-A^{-2}$ emerges from the constraint algebra: the anomaly is proportional to the difference of the quantum and classical evaluation of the loop.    

All this indicates that, even when we do not introduce a quantum group at any stage, and no
dynamical constraint has been imposed yet, amplitudes such as the
value of the quantum dimension (or self linking number of a Wilson
loop in the language of \cite{Witten:1988hc}) $d_{q}=-q-q^{-1}$
and $q=A^2=\exp{i\hbar\lambda/4}$ naturally appear from our
treatment. Recall that the value of $d_q$ together with the
deformed binor identity are the two ingredients for the
combinatorial definition of the Turaev-Viro invariant according to
the formulation of \cite{KL}. This is encouraging as it indicates
that perhaps a strict correspondence between LQG and the
Tuarev-Viro invariant can be established if one appropriately
implements the next step: quantizing and imposing the curvature
constraint (\ref{curvature}). This will be investigated in the
future.

An interesting correspondence between operator ordering and time
was found in \cite{Noui:2004iy} (see also \cite{Perez:2006gja}).
This relationship is expected to be more explicit here. Notice
that even though the canonical quantization is defined on the
2-dimensional manifold $\Sigma$, the non commutativity of the
quantum holonomy, can be encoded in terms of the knotting of paths
as if they would be embedded in a 3-dimensional manifold of
topology $\Sigma\times \R$. If the quantum constraints can be
imposed as in the zero cosmological constant case, we expect the
expectation value of these knots in the physical Hilbert space to
coincide with the ones computed using the covariant methods of
\cite{Witten:1988hc}. This would be an explicit example where
operators defined in the `frozen' timeless formalism of Dirac can
be directly interpreted as space-time processes. Such an example
would be of great conceptual importance showing that the notion of
time and causality can be encoded in the quantum theory defined on
a single space slice.

%{\bf The present work seems to provide further, strong support to the widely shared idea that cosmological constant can be implemented also in the 4d case through a quantum deformation of the Lie algebra. However, it is far from obvious how the path explored here (and more generally the Chern-Simons approach) could be followed also in the four dimensional case, since the non-commutative connection which is so useful in 3d is not available in 4d\footnote{See \cite{Wolf} for a recent application of Chern-Simons theory to the Hamiltonian analysis of the four dimensional case in presence of a cosmological constant}. Nevertheless, in the covariant approach, the cosmological constant has been recently introduced through a quantum deformation of 4d spin foam models \cite{Han}. Our analysis can be regarded as a source of motivation and hope for these models to finally make contact with the canonical approach.}

\section{Acknowledgements}

We would like to thanks useful discussions with Laurent Freidel, Etera Livine and Daniele Oriti.
This work has also benefited from old discussions with 
Rodolfo Gambini and Michael Reisenberger. 
{We are grateful to an unknown referee for helpful comments.} 
AP. would like to thanks the support of 
{\em L'Institut Universitaire de France}.

\appendix

\section{Diagrammatic algebra}\la{app1}

Many results connected to the theory of representation of $SU(2)$
can be more easily stated in a graphical notation introduced by Penrose.
The association of an algebraic meaning to the various diagrams is
subject to many conventions; therefore, here we present ours.

To every single arrow $\begin{array}{c}
\includegraphics{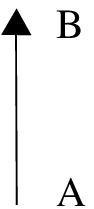}\end{array}$ or $\begin{array}{c}
\includegraphics{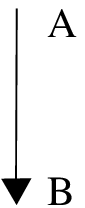}\end{array}$ going from index $A$ to index $B$ associate the symbol $\delta_{B}^{A}$
(note that it does not matter whether the arrow is up- or down-going).

To every symbol $\begin{array}{c}
\includegraphics{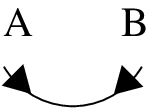}\end{array}$ or $\begin{array}{c}
\includegraphics{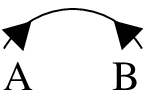}\end{array}$ (ingoing arrows) associate the object $\epsilon^{AB}$ and to every
symbol $\begin{array}{c}
\includegraphics{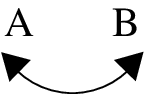}\end{array}$ or $\begin{array}{c}
\includegraphics{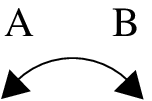}\end{array}$ (outgoing arrows) the object $\epsilon_{AB}$, where $\left(\epsilon_{AB}\right)=\left(\epsilon^{AB}\right)=\left(\begin{array}{cc}
1 & 0\\
0 & -1\end{array}\right)$ (note that it does not matter whether the arc is convex or concave).
Note also that since $\epsilon$ is antisymmetric, $\begin{array}{c}
\includegraphics{epsilon-a-b-in-convex.eps}\end{array}$ is $-\begin{array}{c}
\includegraphics{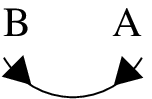}\end{array}$.

It is also important to note that it does not matter whether the strands
are vertical or horizontal, the only important thing being the direction
of the arrows and the reading order of the indices.

With these conventions, it is easy to check that (Penrose's {}``binor
identity'' for $SU(2)$) \be
\begin{array}{c}
\includegraphics{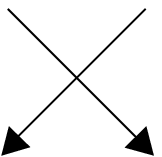}\end{array}=\begin{array}{c}
\includegraphics{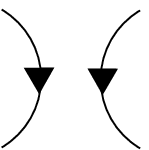}\end{array}-\begin{array}{c}
\includegraphics{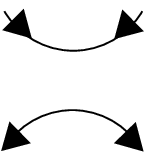}\end{array}\label{bibi}\ee
and that \be
\begin{array}{c}
\includegraphics{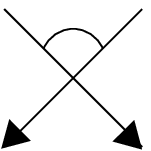}\end{array}=-\frac{1}{4}
\begin{array}{c}
\includegraphics{id-down.eps}\end{array}-\frac{1}{4}\begin{array}{c}
\includegraphics{arc-downdown-downdown.eps}\end{array}.
\label{tauto}
\ee

It is enough to rotate these diagrams in order to get the identities
corresponding to the other three possible choices of arrows.

We also have that \[
\begin{array}{c}
\includegraphics{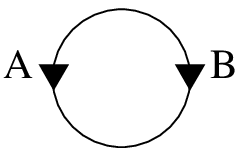}\end{array}=\epsilon_{AB}\epsilon^{AB}=2=\delta_{B}^{A}\delta_{A}^{B}=\begin{array}{c}
\includegraphics{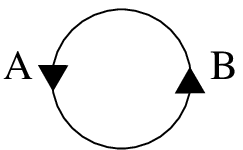}\end{array}\]

and that\[
\begin{array}{c}
\includegraphics{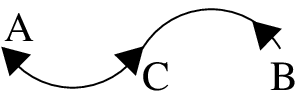}\end{array}=\epsilon_{AC}\epsilon^{CB}=-\delta_{A}^{B}=-\begin{array}{c}
\includegraphics{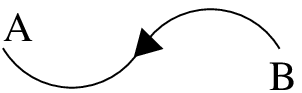}\end{array}.\]

\section{Symmetric ordering}\label{so}

Here we explore the quantization of the quantum holonomy based on
the symmetrized `factor ordering' at the level of (\ref{zz}). This
amounts to replacing the term $\tau_{(i_{k_{1}}}\cdots\tau_{i_{k_{p}})}$
%$\Q(E_{i_{k_{1}}}\cdots E_{i_{k_{p}} })$ 
in Equation
(\ref{esa}) by $\Q_S(E_{i_{k_{1}}}\cdots E_{i_{k_{p}} })=\frac{1}{p!}\sum_{\pi\in S(p)} \tau_{i_{\pi(1)}}\tau_{i_{\pi(2)}}\cdots\tau_{i_{\pi(p)}}$.
%(where $Q_S$means the symmetrization map (\ref{map1q})).

We introduce the Penrose graphical notation\[
\tau^{i_1}\cdots\tau^{i_{p}}
\otimes\tau_{(i_1}\cdots\tau_{i_{p})}=\begin{array}{c}
\includegraphics[height=1.5cm]{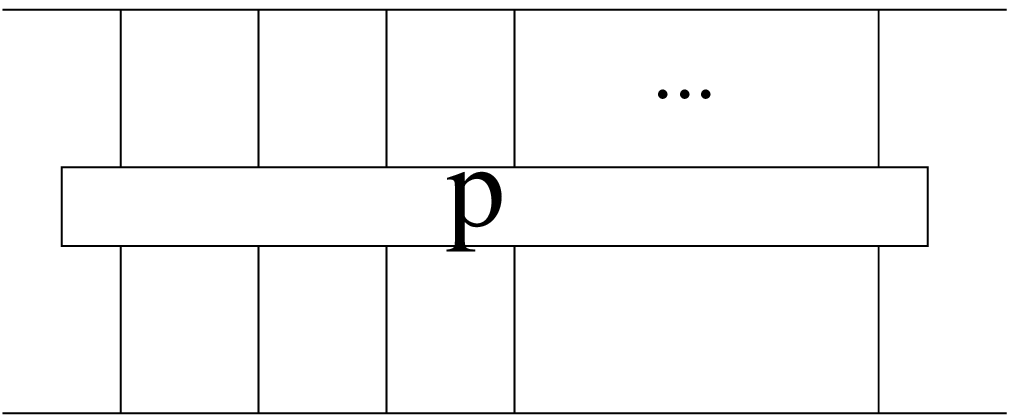}\end{array},
\]
where the vertical lines represent the contraction of the
$i$-indices, the 3-valent nodes denote the $\tau$-matrices, the
horizontal lines represent the contraction of the spinor indices,
i.e., matrix product, and the box in the middle denotes the
symmetrization of the $i$-indices.

Using the fact that $\{\tau^i,\tau^j\}=-2\delta^{ij}$ it is
immediate to proove the following identities: \be
\begin{array}{c}
\includegraphics[height=1.5cm]{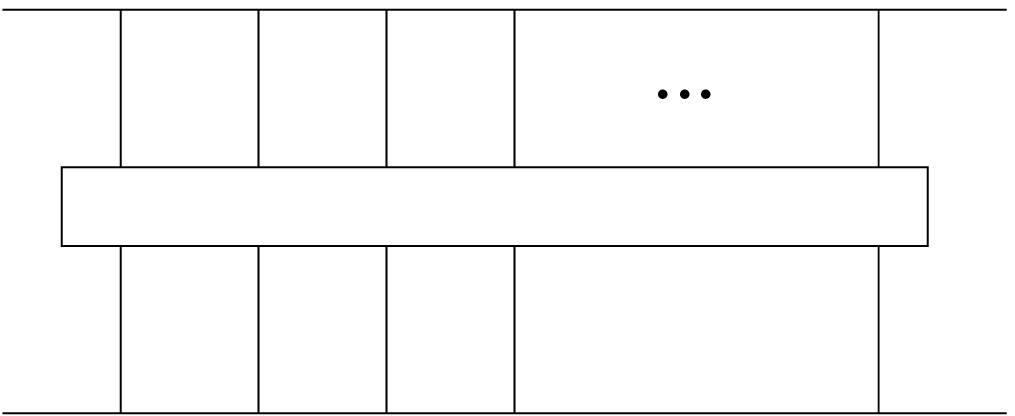}\end{array}=\begin{array}{c}
\includegraphics[height=1.5cm]{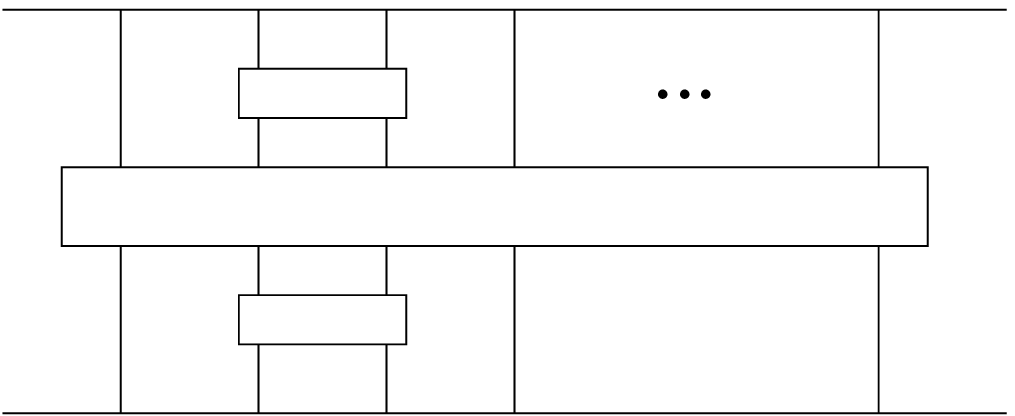}\end{array}=\begin{array}{c}
\includegraphics[height=1.5cm]{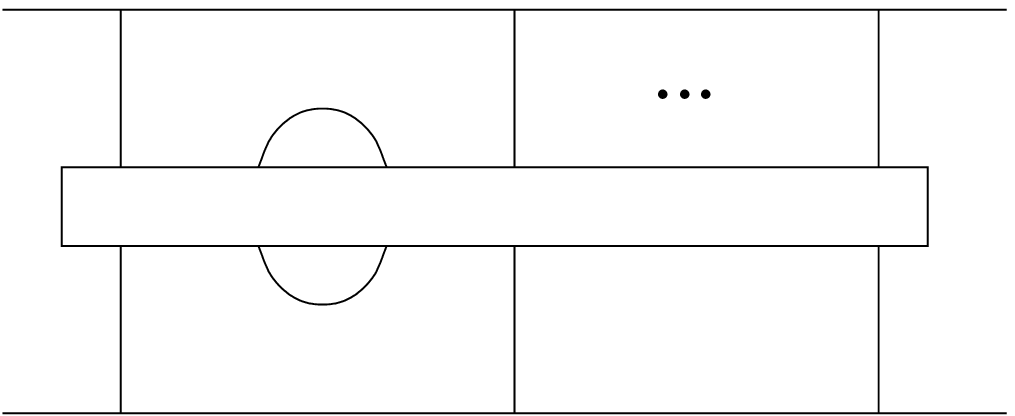}\end{array}
\ee which imply \be
\begin{array}{c}
\includegraphics[height=1.5cm]{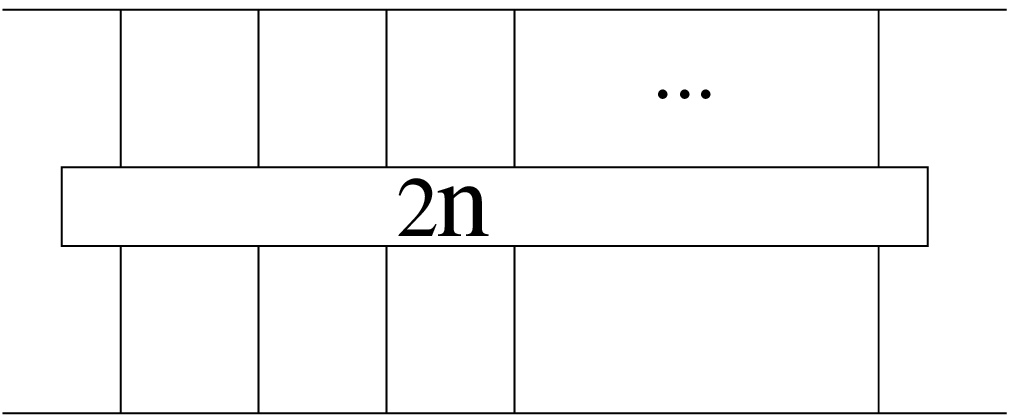}\end{array}=\begin{array}{c}
\includegraphics[height=1.5cm]{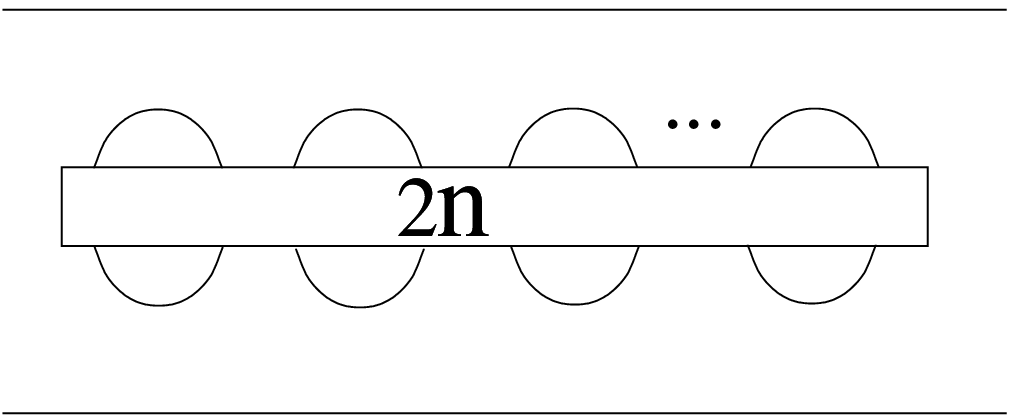}\end{array}=A_{2n}\begin{array}{c}
\includegraphics[height=1.5cm]{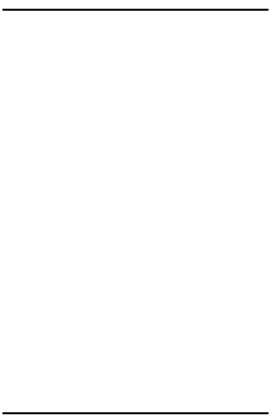}\end{array},
\ee where in the last equality we have introduced the definition
of the coefficient $A_{2n}$, and \be
\begin{array}{c}
\includegraphics[height=1.5cm]{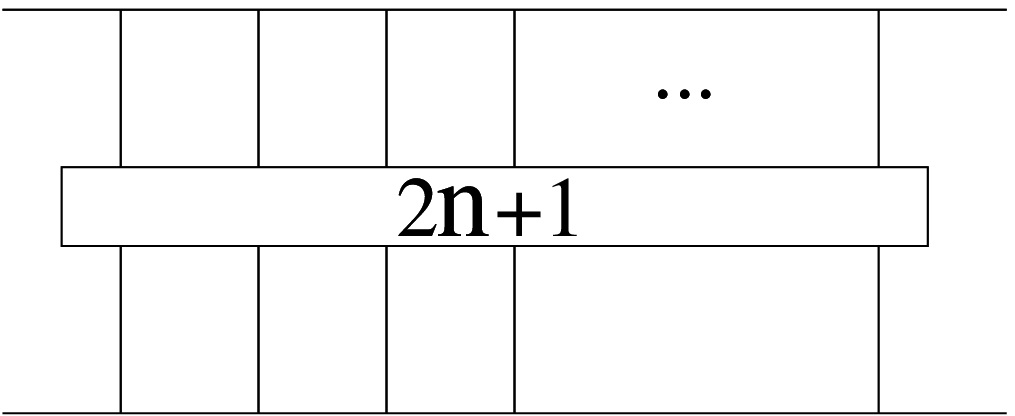}\end{array}=\begin{array}{c}
\includegraphics[height=1.5cm]{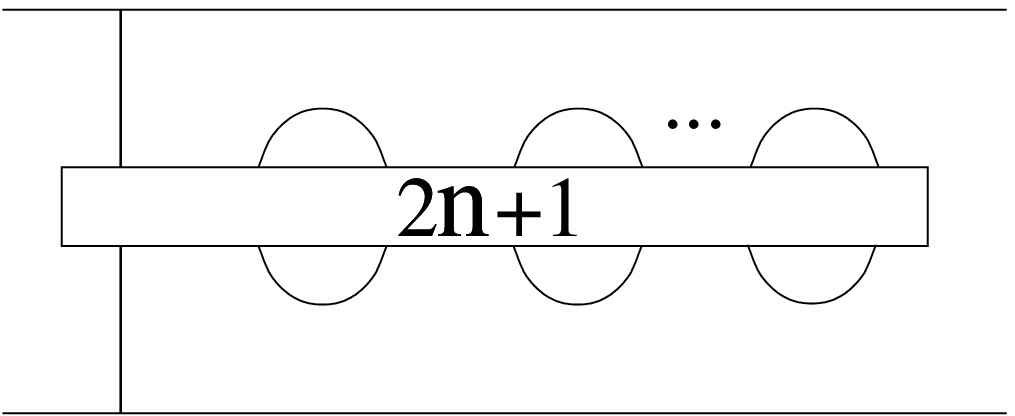}\end{array}=B_{2n+1}\begin{array}{c}
\includegraphics[height=1.5cm]{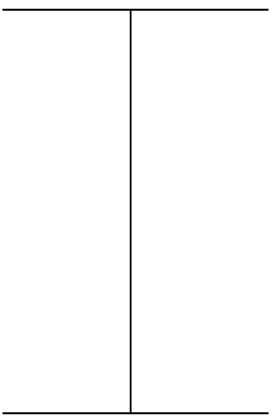}\end{array} ,\label{d3}
\ee where in the last equality we use the fact the the diagram
between the horizontal lines is proportional to the identity in
order to introduce the definition of the coefficient $B_{2n+1}$.
Indeed the previous equations can be written in the standard
tensorial notation as: \be \tau^{i_1}\cdots\tau^{i_{2n}}
\otimes\tau_{(i_1}\cdots\tau_{i_{2n})}=A_{2n} (1\otimes 1),\ee and
\be \tau^{i_1}\cdots\tau^{i_{2n+1}}
\otimes\tau_{(i_1}\cdots\tau_{i_{2n+1})}=B_{2n+1}\
(\tau^i\otimes\tau_i)\ee In order to compute the coefficients
$A_{2n}$ and $B_{2n+1}$ we observe that \ba &&\nonumber
\begin{array}{c}
\includegraphics[width=3cm]{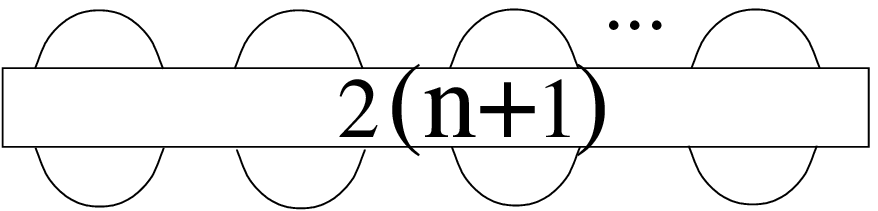}\end{array}=\\
&&\frac{1}{2n+2}\left(\begin{array}{c}
\includegraphics[width=3cm]{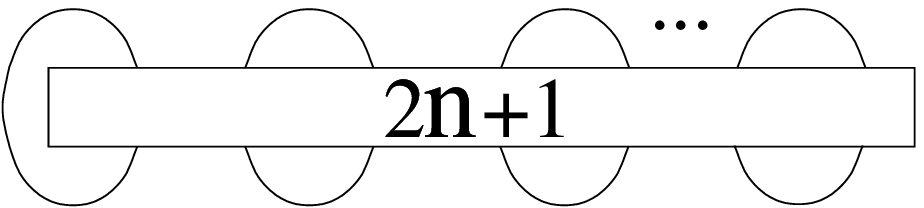}\end{array}+\begin{array}{c}
\includegraphics[width=3cm]{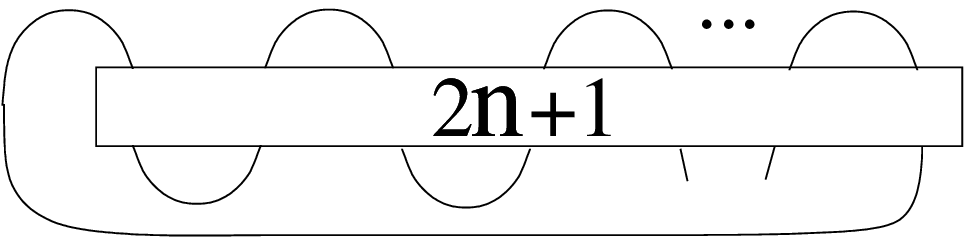}\end{array}+{\rm cyclic \ \
permutations}\right), \ea which is a simple property of symmetric
tensors. But each term on the righthand side is equal to
$B_{2n+1}$ times the trace of the identity (see equation \ref{d3})
Therefore, we have proven that \be 3 B_{2n}=A_{2(n+1)}.\ee  The is
also a simple recursion relation relating the unknown coefficients
which diagrammatically takes the following form:\ba &&\nonumber
\begin{array}{c}
\includegraphics[width=3cm]{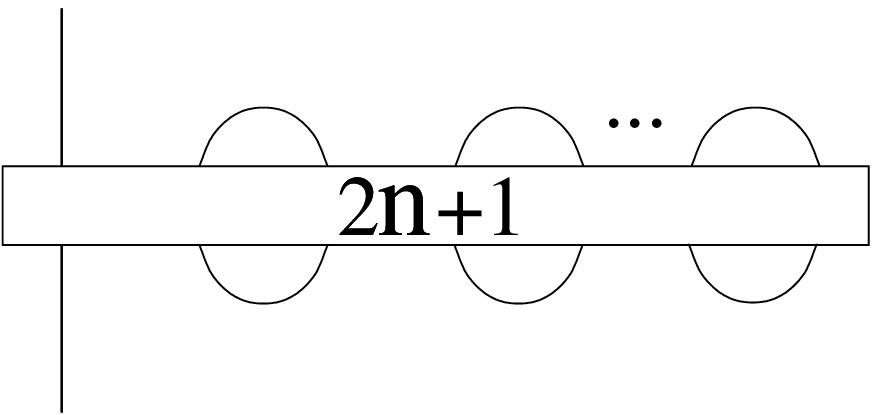}\end{array}=\\ \nonumber
&&\frac{N_0}{2n+1} \begin{array}{c}
\includegraphics[width=3cm]{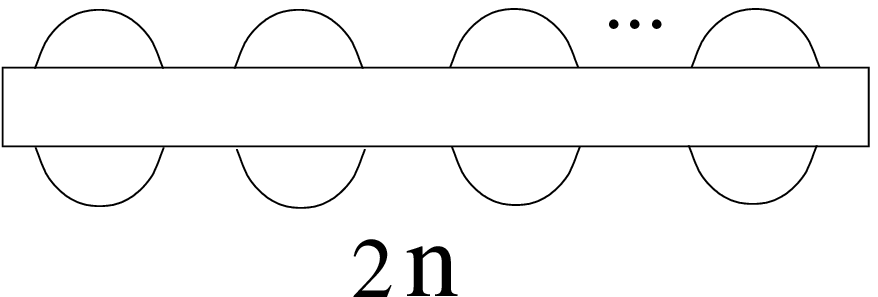}\end{array}  \begin{array}{c}
\includegraphics[height=1.5cm]{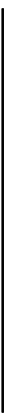}\end{array} + \frac{N_1}{(2n+1)(2n)(2n-1)}\begin{array}{c}
\includegraphics[width=3cm]{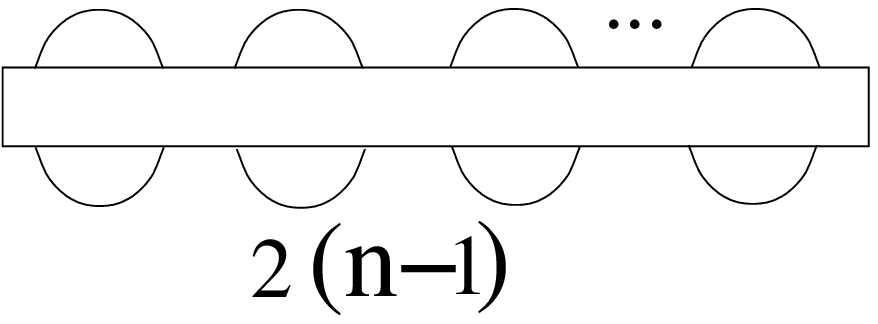}\end{array} \begin{array}{c}
\includegraphics[height=1.5cm]{BN2.eps}\end{array}+\cdots+ N_n \begin{array}{c}
\includegraphics[height=1.5cm]{BN2.eps}\end{array}, \ea
where the factors $N_j$ for $0\le j\le n$ correspond to the number
of ways one can start at the top vertical line go around the
symmetrization box and exit along the bottom vertical line by
`walking' along $j$ upper and $j$ bottom arcs respectively. It is
easy to see that $N_0=1$, $N_1=(2n)^2$ (after entering the box we
have $2n$ choices to enter one of the arcs in the bottom times
$2n$ choices on the top) the general term being
\[N_j=[(2n)(2(n-1))\cdots (2(n-j))]^2=2^{2j}\left[\frac{n!}{(n-j)!}\right]^2\]
The other explicit coefficients in front of each term just come
from the readjustment of the number of permutations. For instance
in the first term $1/(2n+1)$ times the $1/(2n)!$ gives
corresponding to the symmetrization factor on the left
$1/(2n+1)!$. Similarly for the second term we have
$1/((2n+1)(2n)(2n-1))$ times $1/(2(n-1))!$ gives again
$1/(2n+1)!$. The general term being $(2(n-j))!/(2n+1)!$. Putting
all this together we get \be B_{2n+1}=\sum\limits_{j=0}^n \
2^{2j}\, \frac{[2(n-j)]!}{(2n+1)!}\left[\frac{n!}{(n-j)!}\right]^2
A_{2(n-j)} \ee combining the two equations the solution is: \be
A_{2n}=2n+1\ \ \ \ \ B_{2n+1}=\frac{2}{3}n +1 \ee

With this result the symmetrized version of Equation (\ref{esa})
yields \ba && \nonumber \sum_{n\geq
0}\frac{\left(-\mathrm{i}o\hbar
\lambda/4\right)^{p}}{p!}\,\tau^{i_{k_{1}}}\cdots\tau^{i_{k_{p}}}\otimes\tau_{(i_{k_{1}}}\cdots\tau_{i_{k_{p}})}=
\\  \nonumber &&=
\sum_{n\geq 0}\,\frac{\left(-\mathrm{i}o\hbar
\lambda/4\right)^{2n}}{(2n)!}\, (2n+1)
\begin{array}{c}
\includegraphics[width=1cm]{x.eps}\end{array}
+\sum_{n\geq 0}\,\frac{\left(-\mathrm{i}o\hbar
\lambda/4\right)^{2n+1}}{(2n+1)!}\,
\left(\frac{2}{3}n+1\right)\begin{array}{c}
\includegraphics[width=1cm]{x-arc.eps}\end{array}\ea
Finally, choosing an orientation and using eq. (\ref{bibi})-(\ref{tauto}) we
arrive at the result 
\be\la{sim1}
\begin{array}{c}
\includegraphics{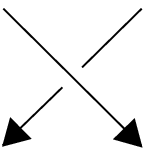}\end{array}=B\begin{array}{c}
\includegraphics{id-down.eps}\end{array}-C\begin{array}{c}
\includegraphics{arc-downdown-downdown.eps}\end{array},\ee
where 
\bee
B(\lambda)=\sin[{\hbar \lambda/4}](\frac{2i}{3}-{\hbar
\lambda/4})+\cos[{\hbar \lambda/4}](1+\frac{i{\hbar
\lambda/4}}{3})
\eee
and
\bee
C(\lambda)=-\sin[{\hbar \lambda/4}](\frac{2i}{3}+{\hbar
\lambda/4})+\cos[{\hbar \lambda/4}](1-\frac{i{\hbar
\lambda/4}}{3}).
\eee
Therefore, considering the totally symmetric map $Q_S$ leads to the wrong result. 
Another possibility consists of trying to improve this map taking into account the action of the flux operators. More precisely, along the lines of Section \ref{nose}, the unambiguous first order term is again given by
\be
\begin{array}{c}
\includegraphics[width=1cm]{x-arc.eps}\end{array}.
\ee 
Then, the second order diagram can be viewed as the result on an
action of the flux operator on the first order diagram. We now apply the symmetrization map $Q_S$ to compute this action, namely
 \be
\begin{array}{c}\psfrag{a}{\!$\va S$}
\includegraphics[width=1cm]{x2.eps}\end{array}=\frac{1}{2}\begin{array}{c}
\includegraphics[width=1cm]{ord2.eps}\end{array}+\frac{1}{2}\begin{array}{c}
\includegraphics[width=1cm]{ord2b.eps}\end{array}=-\frac{1}{4}\begin{array}{c}
\includegraphics[width=1cm]{ord2c.eps}\end{array}=
\frac{3}{16}\begin{array}{c}
\includegraphics[width=1cm]{x.eps}\end{array},
\ee where in the first equation we get two terms coming from on
LIV action and a RIV action, while in the second equality we use
the fact that $\{\tau^i,\tau^j\}=-1/2 \ \delta^{ij}$. Therefore,
the second order diagram is proportional to the order zero
diagram. The proportionality constant is just $1/4$ of the value
of the Casimir in the fundamental representation. The third order
term is consequently proportional to the first order one and so
on. We get in this way the general expression for arbitrary order.
With this prescription the result of the quantum holonomy action now becomes
 \baa
\begin{array}{c}
\includegraphics[width=1cm]{x-quantum.eps}\end{array}=\sum_{n\geq 0}\,\frac{\left(-\mathrm{i}o\hbar
\lambda\right)^{2n}}{(2n)!}\, \left(\frac{3}{16}\right)^n
\begin{array}{c}
\includegraphics[width=1cm]{x.eps}\end{array}
+\sum_{n\geq 0}\,\frac{\left(-\mathrm{i}o\hbar
\lambda\right)^{2n+1}}{(2n+1)!}\, \left(\frac{3}{16}\right)^n\begin{array}{c}
\includegraphics[width=1cm]{x-arc.eps}\end{array},
\eaa
which again, through eq. (\ref{bibi})-(\ref{tauto}), can be written as
 \be\label{sim2}
\begin{array}{c}
\includegraphics{qbinor-down-down.eps}\end{array}=B\begin{array}{c}
\includegraphics{id-down.eps}\end{array}-C\begin{array}{c}
\includegraphics{arc-downdown-downdown.eps}\end{array}\,,\ee
where 
\[B(\lambda)=\cos[\sqrt{3}\hbar
\lambda/4]-\frac{4i}{\sqrt{3}} \sin[{\sqrt{3}\hbar \lambda/4}]\]
and
\[C(\lambda)=\cos[\sqrt{3}\hbar
\lambda/4]+\frac{4i}{\sqrt{3}} \sin[{\sqrt{3}\hbar \lambda/4}].\]
The results (\ref{sim1})-(\ref{sim2}) show how, using some `first guess' ordering to solve the multiple flux operators action ambiguity, one can obtain a series expansion in powers of $\lambda$ which converges and leads to a simple
expression for the crossing, but doesn't reproduce the expected result.


\begin{thebibliography}{10}

\bibitem{Witten:1988hc}
E.~Witten,
  ``Quantum field theory and the Jones polynomial,''
  Commun.\ Math.\ Phys.\  {\bf 121} (1989) 351.
  %%CITATION = CMPHA,121,351;

\newblock  E.~Witten,
  ``(2+1)-Dimensional Gravity as an Exactly Soluble System,''
  Nucl.\ Phys.\  B {\bf 311} (1988) 46.
  %%CITATION = NUPHA,B311,46;%%
  
\newblock  E.~Witten,
  ``Topology Changing Amplitudes in (2+1)-Dimensional Gravity,''
  Nucl.\ Phys.\  {\bf B323}, 113 (1989).
  
  
  
\bibitem{RT}
      N. Reshetikhin. and V.G. Turaev
      "{Invariants of three manifolds via link polynomials and
                        quantum groups}",
      Invent.Math. {\bf 103} (1991) 547.
      

%\cite{Alekseev:1994au}
\bibitem{Alekseev:1994au}
  A.~Y.~Alekseev, H.~Grosse and V.~Schomerus,
  ``Combinatorial quantization of the Hamiltonian Chern-Simons theory. 2,''
  Commun.\ Math.\ Phys.\  {\bf 174} (1995) 561.
  [arXiv:hep-th/9408097].
  %%CITATION = CMPHA,174,561;%%
 \newblock A.~Y.~Alekseev, H.~Grosse and V.~Schomerus,
  ``Combinatorial quantization of the Hamiltonian Chern-Simons theory,''
  Commun.\ Math.\ Phys.\  {\bf 172} (1995) 317.
  [arXiv:hep-th/9403066].
  %%CITATION = CMPHA,172,317;%%
\newblock  V.~V.~Fock and A.~A.~Rosly,
  %``Poisson structure on moduli of flat connections on Riemann surfaces and
  %r-matrix,''
  Am.\ Math.\ Soc.\ Transl.\  {\bf 191} (1999) 67.
  [arXiv:math/9802054].
  %%CITATION = AMTLA,191,67;%%


\bibitem{BNR}
 E. Buffenoir, K. Noui and P. Roche,
      "{Hamiltonian quantization of Chern-Simons theory with
                        SL(2,C) group}",
      Class.Quant.Grav.
      {\bf 19} (2002) 4953.
      [arXiv:hep-th/0202121].
   


\bibitem{Cat}
C. Meusburger and K. Noui,
     "{Combinatorial quantisation of the Euclidean torus universe}",
      Nucl.Phys. {\bf B841} (2010) 463.
      [arXiv:1007.4615].


%\cite{Turaev:1992hq}
\bibitem{Turaev:1992hq}
  V.~G.~Turaev and O.~Y.~Viro,
  ``State sum invariants of 3 manifolds and quantum 6j symbols,''
  Topology {\bf 31} (1992) 865.
  %%CITATION = TPLGA,31,865;%%
  

%\cite{Mizoguchi:1991hk}
\bibitem{Mizoguchi:1991hk}
  S.~Mizoguchi and T.~Tada,
  ``Three-dimensional gravity from the Turaev-Viro invariant,''
  Phys.\ Rev.\ Lett.\  {\bf 68} (1992) 1795.
  [arXiv:hep-th/9110057].
  %%CITATION = PRLTA,68,1795;%%

\newblock Y. Taylor and C. Woodward, ``6j symbols for $U_q(sl_2)$ and non-Euclidean tetrahedra,''
  [arXiv:math/0305113].


\bibitem{Noui:2004iy}
  K.~Noui and A.~Perez,
  ``Three dimensional loop quantum gravity: Physical scalar product and  spin
  foam models,''
  Class.\ Quant.\ Grav.\  {\bf 22} (2005) 1739.
  [arXiv:gr-qc/0402110].
  %%CITATION = CQGRD,22,1739;%%

\bibitem{Bonzom-Freidel}
 V.~Bonzom, L.~Freidel,
  ``The Hamiltonian constraint in 3d Riemannian loop quantum gravity,''
  [arXiv:1101.3524 [gr-qc]].

 

%\cite{Ashtekar:1998ak}
\bibitem{zapa}
  A.~Ashtekar, A.~Corichi and J.~A.~Zapata,
  ``Quantum theory of geometry. III: Non-commutativity of Riemannian
  structures,''
  Class.\ Quant.\ Grav.\  {\bf 15} (1998) 2955.
  [arXiv:gr-qc/9806041].
  %%CITATION = CQGRD,15,2955;%%



\bibitem{lqg}
T. Thiemann, ``Modern Canonical Quantum General Relativity''
Cambridge, UK: Cambridge Univ. Pr. (2007) 819 p;
  [arXiv:gr-qc/0110034]. 
  
  \newblock C. Rovelli,
    `` Quantum gravity,'' Cambridge, UK: Univ. Pr. (2004) 455 p.

\newblock A.~Ashtekar and J.~Lewandowski,
  ``Background independent quantum gravity: A status report,''
  Class.\ Quant.\ Grav.\  {\bf 21}, R53 (2004).
  [arXiv:gr-qc/0404018]. 
  
  \newblock A.~Perez,
  ``Introduction to loop quantum gravity and spin foams,''
Proceedings of the International Conference on Fundamental
Interactions, Domingos Martins, Brazil, (2004).
[arXiv:gr-qc/0409061].

%\cite{Lewandowski:2005jk}
\bibitem{Lewandowski:2005jk}
  J.~Lewandowski, A.~Okolow, H.~Sahlmann and T.~Thiemann,
  ``Uniqueness of diffeomorphism invariant states on holonomy-flux  algebras,''
  Commun.\ Math.\ Phys.\  {\bf 267}, 703-733 (2006).
  [gr-qc/0504147].
  %%CITATION = GR-QC/0504147;%%

\bibitem{Rovelli:1995ac}
  C.~Rovelli and L.~Smolin,
  ``Spin networks and quantum gravity,''
  Phys.\ Rev.\  D {\bf 52} (1995) 5743.
  [arXiv:gr-qc/9505006].
  %%CITATION = PHRVA,D52,5743;%%

\bibitem{3}
J.\ C.~Baez, ``Generalized measures in gauge theory'', Lett.\
Math.\ Phys.\ {\bf 31} 213, (1994).
  [arXiv:/hep-th/9310201].

\newblock J.\ C.~Baez, ``Diffeomorphism-invariant generalized measures on
the space of connections modulo gauge transformations,'' in {\sl
Proceedings of the Conference on Quantum Topology}, ed.\ D.\ N.
Yetter, World Scientific Press, Singapore, 1994.
  [arXiv:hep-th/9305045].
 
 \newblock A.~Ashtekar and J.~Lewandowski,
  ``Projective techniques and functional integration for gauge theories,''
  J.\ Math.\ Phys.\  {\bf 36}, 2170 (1995).
  [arXiv:gr-qc/9411046]. 
  
  \newblock A.~Ashtekar and J.~Lewandowski, ``Differential geometry on the space of connections via graphs and projective
  limits,'' J.\ Geom.\ Phys.\  {\bf 17}, 191 (1995).
  [arXiv:hep-th/9412073].

\bibitem{jaco}
  T.~Jacobson and L.~Smolin,
  ``Nonperturbative Quantum Geometries,''
  Nucl.\ Phys.\  B {\bf 299} (1988) 295.
  %%CITATION = NUPHA,B299,295;%%


\bibitem{KL}
  L.~H.~Kauffman,
  ``Knots and physics,''
{\it  Singapore, Singapore: World Scientific (1991) 538 p. (Series
on knots and everything, 1)}. 

\newblock L.H.~Kauffmann and S. Lins,
''Temperley-Lieb Recoupling Theory and Invariants of 3-Manifolds''
Annals of Mathematical Studies, Princeton Univ. Press. 
%\cite{Perez:2006gja}

\bibitem{Perez:2006gja}
  A.~Perez,
  ``The spin-foam-representation of loop quantum gravity,''
  [arXiv:gr-qc/0601095]. To appear in
  "Towards quantum gravity", Ed. Daniele Oriti, Cambridge University Press.
  %%CITATION = GR-QC/0601095;%%

\bibitem{spinfoams}
  A.~Perez,
  ``Spin foam models for quantum gravity,''
  Class.\ Quant.\ Grav.\  {\bf 20}, R43 (2003).
  [arXiv:gr-qc/0301113]. 
  
  \newblock D.~Oriti,
  ``Spacetime geometry from algebra: Spin foam models for non-perturbative
  quantum gravity,''
  Rept.\ Prog.\ Phys.\  {\bf 64}, 1489 (2001).
  [arXiv:gr-qc/0106091].
  
  \newblock J.~C.~Baez,
  ``An introduction to spin foam models of $BF$ theory and quantum gravity,''
  Lect.\ Notes Phys.\  {\bf 543}, 25 (2000).
  [arXiv:gr-qc/9905087].  
  
  \newblock J.~C.~Baez,
  ``Spin foam models,''
  Class.\ Quant.\ Grav.\  {\bf 15}, 1827 (1998).
  [arXiv:gr-qc/9709052].
  

\bibitem{SF-LQG}
 E.~Alesci, K.~Noui, F.~Sardelli,
  ``Spin-Foam Models and the Physical Scalar Product,''
  Phys.\ Rev.\  {\bf D78}, 104009 (2008).
  [arXiv:0807.3561 [gr-qc]].

 \newblock E.~Alesci, C.~Rovelli,
  ``A Regularization of the hamiltonian constraint compatible with the spinfoam dynamics,''
  Phys.\ Rev.\  {\bf D82}, 044007 (2010).
  [arXiv:1005.0817 [gr-qc]].
  
  \newblock V.~Bonzom,
  ``Spin foam models and the Wheeler-DeWitt equation for the quantum 4-simplex,''
  Phys.\ Rev.\  {\bf D84}, 024009 (2011).
  [arXiv:1101.1615 [gr-qc]].


  \bibitem{Anomaly}
  A.~Perez, D.~Pranzetti,
  ``On the regularization of the constraints algebra of Quantum Gravity in 2+1 dimensions with non-vanishing cosmological constant,''
  Class.\ Quant.\ Grav.\  {\bf 27}, 145009 (2010).
  [arXiv:gr-qc/10013292].
  
  \bibitem{Ansatz}
  D.~Pranzetti,
  ``2+1 gravity with positive cosmological constant in LQG: a proposal for the physical state,''
  [arXiv:gr-qc/11015585].

\bibitem{Duflo} M. Duflo, ``Op\'erateurs diff\'erentiels bi-invariants sur un groupe de Lie,'' Ann. scient. Ecole Norm. Sup. {\bf10}, 265288 (1977).

\bibitem{Majid}
  L.~Freidel, S.~Majid,
  ``Noncommutative harmonic analysis, sampling theory and the Duflo map in 2+1 quantum gravity,''
  Class.\ Quant.\ Grav.\  {\bf 25}, 045006 (2008).
  [hep-th/0601004].

\bibitem{Sahlmann}
  H.~Sahlmann, T.~Thiemann,
  ``Chern-Simons theory, Stokes' Theorem, and the Duflo map,''
  J.\ Geom.\ Phys.\  {\bf 61}, 1104-1121 (2011).
  [arXiv:1101.1690 [gr-qc]].
  
  \bibitem{Alekseev}
  A.~Alekseev, A.~P.~Polychronakos, M.~Smedback,
  ``On area and entropy of a black hole,''
  Phys.\ Lett.\  {\bf B574}, 296-300 (2003).
  [hep-th/0004036].

\bibitem{alex}
  A.~Corichi,
  ``Comments on area spectra in loop quantum gravity,''
  Rev.\ Mex.\ Fis.\  {\bf 50} (2005) 549.
  [arXiv:gr-qc/0402064].
  %%CITATION = RMXFA,50,549;%%

\bibitem{Wolf} W. Wieland, ``Complex Ashtekar variables, the Kodama state and
spinfoam gravity,'' [arXiv:gr-qc/1105.2330].

\bibitem{Han}
M.~Han,
  ``4-dimensional Spin-foam Model with Quantum Lorentz Group,''
  J.\ Math.\ Phys.\  {\bf 52}, 072501 (2011).
  [arXiv:1012.4216 [gr-qc]].

\newblock  M.~Han,
  ``Cosmological Constant in LQG Vertex Amplitude,''
  [arXiv:1105.2212 [gr-qc]].
  
\newblock  W.~J.~Fairbairn, C.~Meusburger,
  ``Quantum deformation of two four-dimensional spin foam models,''
  [arXiv:1012.4784 [gr-qc]].

\end{thebibliography}
\end{document}